\documentclass[twocolumn,twocolappendix]{openjournal}
\usepackage{lastpage}
\usepackage{xcolor}
\begin{document}
\title{Is the universe static?}
\email{dcrawfrd@bigpond.net.au}
\author{David F. Crawford}
\affil{Astronomical Society of Australia,
44 Market St., Naremburn NSW,  2065,
Australia}
\begin{abstract}
A fundamental property of an expanding universe is that any  time dependent characteristic of distant objects must appear to scale by the factor $(1+z$). This is called time dilation.  Light curves of type Ia supernovae and the duration of Gamma-Ray Bursts (GRB) are the only observations that can directly measure time dilation over a wide range of redshifts. An analysis of raw observations of 2,333 type Ia supernovae light-curves shows that their widths, relative to a standard template, have a power-law exponent as a function of ${(1+z)}$, of (0.083 +/- 0.024) which is consistent with no time dilation and inconsistent with standard time dilation.  In addition, it is shown that the standard method for calibrating the type Ia supernovae light curves (SALT2) is flawed, which explains why this lack of time dilation has not been previously observed.
\par
Nearby observations show that the peak absolute magnitude of type Ia supernovae is also constant. Here it is shown that the peak absolute magnitude is independent of redshift if a static universe cosmology, Curvature Cosmology,  is used to provide the distance moduli. Furthermore it is explained  why the  modified $\Lambda$-CDM model provides similar results.
\par
Analysis of the duration of GRB shows that they are  consistent with no time dilation and have no support for standard time dilation.
Consequently, this paper argues for a fundamental change from the current paradigm of an expanding universe to one for a static universe. Some of the major consequences of Curvature Cosmology  are listed.
\par
\end{abstract}
\par
\section{\bf{INTRODUCTION}}
\label{s1}
Nearby type Ia supernovae are well known to have essentially identical light curves that make  excellent cosmological probes. It is argued in Section~\ref{s2} that the only characteristics of the light curve that changes with redshift are the scaling parameters of peak luminosity and  width. In particular  the width which must vary with redshift in exactly the same way as time dilation and  the peak absolute magnitude should be independent of redshift. Investigation of the variation of the peak magnitude with redshift requires a cosmological model to provide the distance moduli. This is done by using Curvature Cosmology  \citep{Crawford10} which will be described later (Section~\ref{s8}).
\par
\subparagraph{Time dilation}
It is convenient to assume that the width dependence on redshift can be described by a parameter $\alpha$ such that the redshift dependence of time dilation is equal to $(1+z)^\alpha$ and then to estimate $\alpha$ from the observations as a test of time dilation. For all current expansion models $\alpha$ must be one and for static models, it must zero. Any significant difference of $\alpha$ from either of these two values could only be explained by  a completely new cosmology.
\par
\subparagraph{Photon energy}
In Section~\ref{s3} it is argued that in quantum mechanics the apparent wavelength of a photon is a measurement of its energy and as a consequence  its redshift may be due to any process that causes a loss of energy. Thus in quantum mechanics, the rigid nexus between the shift of spectral lines and  other time variations is broken.
\par
\subparagraph{Intrinsic widths}
The observational evidence for standard time dilation has a long history with notable papers being \citet{Goldhaber01} and \citet{Blondin08}. The observed width of any light curve from a distant object is the product of an intrinsic width, with a cosmological width due to  time dilation.  If the observed wavelength is $\lambda$ then the intrinsic wavelength is $\epsilon=\lambda/(1+z)$  which is  shorter than the observed wavelength. Since many of the intrinsic wavelengths are beyond the visible range, their intrinsic widths cannot be easily determined from nearby supernovae. A suitable method of solving this problem is to  generate a reference template from all the supernovae light curves that provides a complete light curve for each intrinsic wavelength, and then to use these templates to accurately calibrate the observations by eliminating any intrinsic effects.
\par
\subparagraph{Salt2}
The SALT2 method \citep{guy07,Guy10} determines these intrinsic templates by combining a large number of observations over a wide range of redshifts and has been used by \citet{Betoule14}, \citet{Conley11}, \citet{Foley18}, \citet{Scolnic17} and \citet{Jones18}. In other words, the reference template is the average of the  light curves from many supernovae as a function of intrinsic wavelength. Consequently the measurement of the intrinsic width at a particular intrinsic wavelength can come from many supernovae with different redshifts.
\par
It has been first shown by \citet{Crawford17} and here in Section~\ref{s4} that there is a fundamental problem with the SALT2 analysis in that a systematic variation in width  as a function of redshift is included in the template as a systematic variation of the width as a function of intrinsic wavelength. The SALT2 calibration process is very good at removing the intrinsic variations but at the same time, it removes systematic redshift variations such as time dilation. Thus supernovae light curves that have been calibrated by SALT2, or a similar method, have  all the cosmological information that is a power-law function of redshift removed.
\par
\subparagraph{Type Ia supernova light curve widths}
A major part of this paper (Section~\ref{s5}) is an examination of the raw observations for 2,333 supernovae  to investigate how the widths of their light curves vary with redshift. This was done separately for each filter so that observations of each supernova can provide up to five distinct and unrelated values for the width of the observed light curve.
\par
The first step of the  analysis  is to determine the intrinsic width as a function of intrinsic wavelength and the second step is to use the intrinsic width to remove the intrinsic component from the raw observations in order to obtain a cosmological width that is entirely due to cosmological processes.
\par
\subparagraph{Type Ia supernovae absolute magnitudes}
It has also been observed that nearby type Ia supernovae have peak absolute magnitudes that are independent of redshift. The aim here is to extend this analysis of peak absolute magnitudes to higher redshifts. However there is a major problem in that the estimation of the absolute magnitude  requires a distance modulus to convert the apparent flux density to an absolute magnitude and this requires the use of a cosmological model.  This analysis is done for two different models: the standard $\Lambda$-CDM model (summarised in Appendix~\ref{sa2}) and a static model, Curvature Cosmology (Section~\ref{s8}).
\par
The first step of the analysis is to get the average absolute intrinsic magnitude as a function of intrinsic wavelength. Then to remove these intrinsic effects from the observations to provide cosmological peak absolute magnitudes for each filter and each supernova.  This is done, separately, for both cosmological models.
\par
\subparagraph{Gamma- Ray Bursts}
Gamma-Ray Bursts (GRB) are the only other observations that can provide direct
measurements of time dilation over a wide range of redshifts. The analysis used here in Section~\ref{s6} investigates the observed durations of the bursts as a function of redshift and again finds that they are consistent with no time dilation.
\par
\subparagraph{Curvature Cosmology}
Following the analysis a brief description of Curvature Cosmology is provided (Section~\ref{s8}) as well  as the description of the cosmological consequences of  Curvature Cosmology (Section~\ref{s8.10}).
\par
\subparagraph{Basic assumptions}
It is assumed for this analysis  that the intrinsic properties of the type Ia supernovae light curves and the GRB burst durations are the same at all redshifts. In other words, we are assuming that there is no evolution.  Part of this assumption for type Ia supernovae is that minor differences in the subtypes and effects of the host galaxy do not have a significant dependence on redshift. Hence their main effect is to increase the background noise. Furthermore it is assumed that the wavelength dependence of a filter can be replaced by a single value at the effective wavelength of the filter.
\vspace{0.3cm}
\section{\bf{REDSHIFTS AND TIME DILATION}}\label{s3}
The Hubble redshift law states  that distant objects appear, on average,  to have an apparent velocity of recession that is proportional to their distance. Since this is consistent with models in General Relativity that have universal expansion, such expanding models have become the standard cosmological paradigm.  Classically, this redshift was obvious because in these models spectral lines are shifted in wavelength exactly like any other time-dependent phenomena.
\par
However quantum mechanics tells us that light is transmitted by photons whose effective wavelength is determined from  their momentum  by the de Broglie equation $\lambda=hc/E$ where $E$ is their energy and $\lambda$ is their effective wavelength. Thus their effective wavelength is simply a measurement of their energy and is not a proper wavelength in the classical sense. Nevertheless it does describe how photons can be diffracted and their energy measured by an interferometer. The Doppler effect and the universal expansion are explained by an actual loss (or gain) of photon energy.  A consequence is that redshifts may be due to any process that causes a loss of photon energy. Thus in quantum mechanics, the rigid nexus between the shifts in wavelength of spectral lines and  other time variations is broken.
\section{\bf{COSMOLOGICAL CHARACTERISTICS OF LIGHT CURVES}}\label{s2}
Let us assume that the intrinsic radiation characteristics of type Ia supernovae  are  independent of redshift and that in an expanding universe  the rate of universal expansion is constant for the duration of the light curve. Since cosmology only controls the transmission of the light, it follows that the shape of the received light-curve must be the same as the shape of the intrinsic light curve but with   different scale factors. In other words, the cosmology can only change the peak flux density and the width of the light curve. Consequently, all of the cosmological information is contained in the dependence of these two variables  with redshift. Thus, it is only necessary to measure these two scaling parameters in order to investigate the cosmology of supernovae light curves. Furthermore, we only need determine these two parameters for intrinsic light curves in order to distinguish them from cosmological effects.
\par
Since the observed light curve is  the intrinsic light curve multiplied by any time dilation, then the observed width  is the product of the intrinsic width and the time dilation width. We assume  that the time dilation  width has the  power law of ${(1+z)}^\alpha$, where $\alpha$ is the exponent and has a value of one for standard time dilation.  By definition, the redshift, $z$, is defined by $\epsilon=\lambda/{(1+z)}$, where $\lambda$ is the observed wavelength and  $\epsilon$ is the intrinsic (emitted)  wavelength.
\par
The observed wavelength is determined by the filter used and the redshift of the supernova is usually measured from the observed wavelength shift of emission or absorption lines. The intrinsic width is a function of the intrinsic wavelength and we assume for part of this work that it has the  power law $\epsilon^\beta$, where $\beta$ is determined by observations. Although the intrinsic light curve width almost certainly has a more complicated function of wavelength, it is only that  part of it that can be described by this power law that will enable it to be confused with time dilation. Hence the model used here is that the observed width, $w(\lambda)$ is
\begin{equation}
\label{f2}
w(\lambda) = {(1+z)}^{\alpha}\epsilon^{\beta},
\end{equation}
where the   reference width is one.  Substituting for $\epsilon$ provides the more informative equation
\begin{equation}
\label{f3}
w(\lambda) = \lambda^{\beta} {(1+z)}^{(\alpha-\beta)}.
\end{equation}
This shows the close relationship between intrinsic and cosmological widths.
\par
Note that there is a common intrinsic variation in either light curve width or absolute peak magnitude for all filters. This intrinsic curve replaces all the k-corrections, colour corrections, and similar methods used to correct observations to standard filters or redshifts.
\section{\bf{INTRINSIC DEPENDENCE OF LIGHT CURVES}}
\label{s4}
Observations of local type Ia supernovae show that the emission from the expanding gas cloud is multicoloured and the intensity is a function of both wavelength and time. A major practical problem is that the emitted wavelengths are often much shorter than the observed wavelengths and since the shape and size of the intrinsic light curve is a function of the wavelength the analysis of observations requires that this intrinsic dependence is known. For high redshift supernovae, many of the emitted wavelengths are outside the visual range, which means that we cannot, in general,  use nearby supernovae to obtain the required calibrations.
\par
An ingenious solution, exemplified by the SALT2 method \citep{guy07,Guy10}, is to determine the calibration spectra from averaging  the light curves of many supernovae at many different redshifts. Because the only observations available are from filters that cover a large wavelength range this is a difficult process. This and similar methods carefully deconstruct the average light curve, as a function of intrinsic wavelengths from a large number of observations,  and then generate a light-curve template for each intrinsic wavelength. Thus the light curve for any particular intrinsic wavelength will have contributions from supernovae at many observed wavelengths.
\subsection{A flaw in the SALT2 method}
\label{s4.1}
However, there is a problem first described by \citet{Crawford17} with the SALT2  method of determining the characteristics of the intrinsic light curve. Let $w(\lambda)$ be the observed width at, $\lambda$, and let $W(\epsilon)$ be the width at the intrinsic (rest frame) wavelength $\epsilon$. (The use of $w$ and $W$ was chosen to mimic the familiar use of $m$ and $M$ for magnitudes.) Similarly, let $f(\lambda,z)$ and $F(\epsilon,z)$ be the observed and emitted flux densities.
\par
Equation~\ref{f3}  shows that there is a close correspondence between a systematic variation in width with intrinsic wavelength and time dilation.  Although this is interesting, the extension to a wide range of wavelengths needs a more refined analysis.
\par
The supernovae observations typically consist of the flux-density measure in filters that essentially cover the visual wavelengths. Since each filter has a filter gain function, $g(\lambda)$ which is the fraction of power transmitted per unit wavelength, then the flux density observed by a particular filter at the  wavelength $\lambda$ is given by
\begin{equation}
\label{e1}
 f(\lambda,z) \propto \int g(\lambda^*)F(\epsilon,z)\,d\lambda^*,
\end{equation}
where $\lambda^*$ is a dummy integration variable. Now the  width of the light curve can be determined by measuring the difference between the two half peak flux density points. To the first order, this process is a linear function of the flux densities and will certainly be sufficiently linear in the last step of an iterative process of measuring the width.  Thus we can apply Eq.~\ref{e1}  to the widths to get
\begin{equation}
\label{e2}
w(\lambda,z) = \int g(\lambda^*)W(\epsilon)\,d\lambda^*.
\end{equation}
Now suppose the intrinsic light curves have a power-law wavelength dependence so that $W(\epsilon)=\epsilon^{\beta}$ where $\beta$ is a constant.
Then including this power law in Eq.~\ref{e2} gives
\begin{equation}
\label{e3}
w(\lambda,z) = \int g(\lambda^*)\left(\frac{\lambda^*}{1+z}\right)^\beta \,d\lambda^*.
\end{equation}
Since the $(1+z)$ term is independent of $\lambda^*$ it can be taken outside of the integral to get
\begin{equation}
\label{e4}
w(\lambda,z) = A(1+z)^{-\beta},
\end{equation}
where
\begin{equation}
\label{e5}
A = \int g(\lambda)\lambda^\beta\, d\lambda.
\end{equation}
Clearly, $A$ is only a function of $\beta$ and the filter characteristics and it is the same for all observations with this filter. Consequently, if the intrinsic widths have a power-law dependence on wavelength, proportional to $\epsilon^\beta$ and from Eq.~\ref{e4}, this will be seen as a power-law dependence of the observed widths that is proportional to $(1+z)^{-\beta}$.
\par
Conversely, if there is no intrinsic variation of the widths of the light curve with wavelength but there is a  time dilation  with exponent $\alpha$, then the derived intrinsic wavelength dependence in the SALT2 template from multiple supernovae will have a power-law dependence with $\beta=-\alpha$.
\par
In practice, this means that during the generation of a reference spectrum, any observed time dilation is recorded in the templates as an intrinsic  wavelength dependence. When this is used to calibrate new observations that (by definition) have the same time dilation, this redshift dependence in the observations will be cancelled by the wavelength dependence in the template and the calibrated widths will be independent of redshift.
\par
Consequently,  if the SALT2 or similar calibration method is used then any cosmological information that was in the calibration observations in the form of a power law of $(1+z)$, will be removed from subsequent analyses. Simply put, the SALT2 calibration  removes all power laws as function of $(1+z)$, whether artificial or genuine, leaving the calibrated light curve without any of this power-law information and therefore without any cosmological information.
\subsection{SALT2 template analysis}\label{s4.2}
\begin{figure}
\includegraphics[width=\columnwidth]{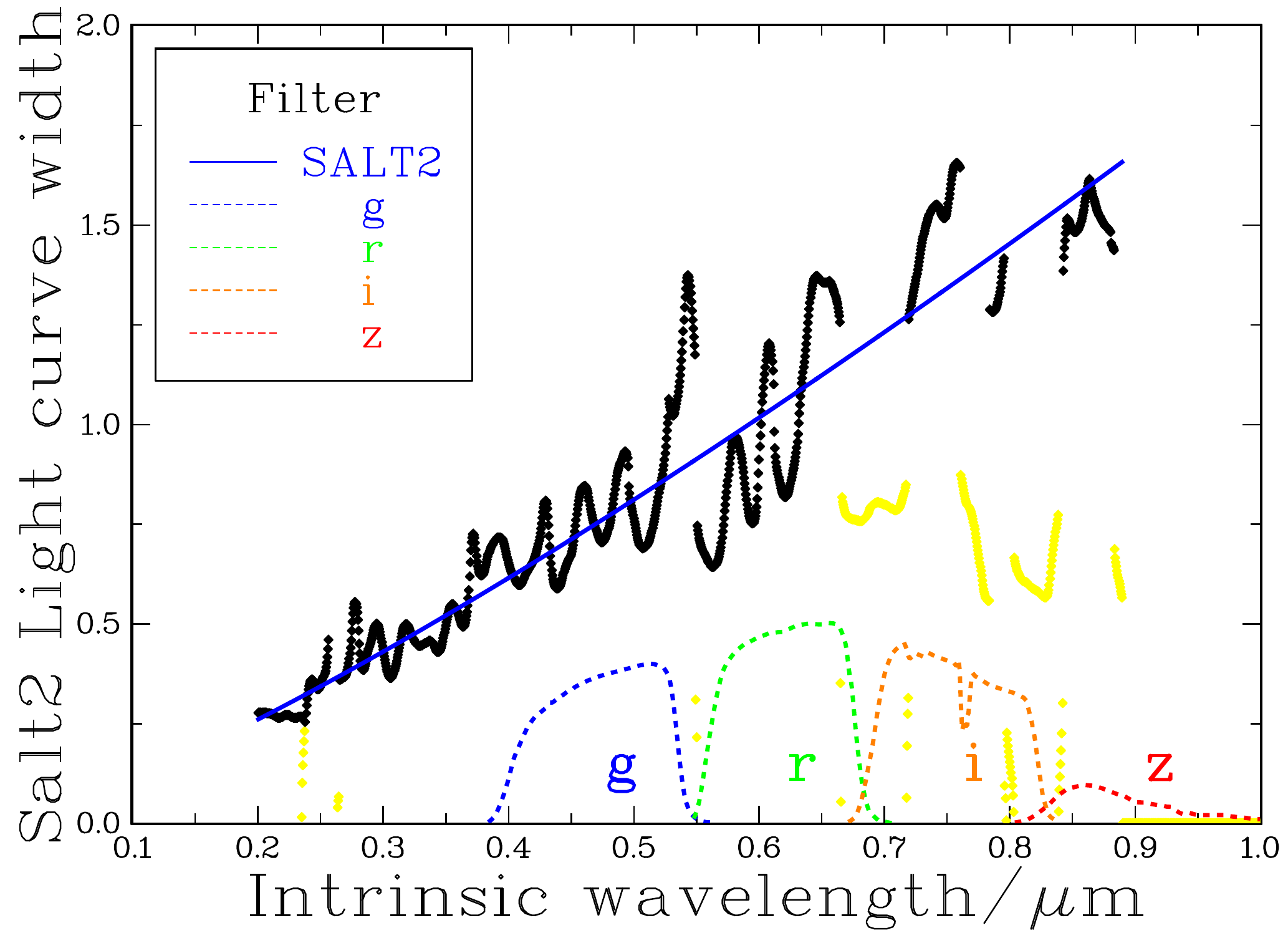}
\caption{\label{fig1} A plot, in black,  of the relative widths of the light curves for the SALT2 templates as a function of intrinsic wavelength. The blue line is the best fit power law with an  exponent $1.240\pm0.014$. Yellow points are assumed to be invalid. For comparison some of the filter response curves for nearby supernovae are also shown.}
\end{figure}

In order to remove the expected time dilation, the first step of the standard SALT2 analysis is to divide all the epoch differences by ${(1+z)}$. If there is no time dilation, this will produce an effective time dilation of ${(1+z)}^{-1}$. Figure~\ref{fig1} shows the relative width (in black and yellow points) for each intrinsic wavelength  of the light curves in the SALT2 template (cf. Appendix~\ref{a1}). Since there are clearly problems with some of the widths, shown in yellow, the analysis was confined to the black  points. The explanation for the bad  widths is unknown but one contributing factor could be poor data for wavelengths between the filters. Shown in Figure~\ref{fig1} are some filter response curves for the nearby supernovae where this effect would be most pronounced.
\par
The blue line is the best power-law fit of the black points and has an exponent of  $\beta=1.240\pm0.011$. Then allowing for the initial division of the epoch differences by ${(1+z)}$ either  there is no time dilation ($\alpha=0$) and an intrinsic dependence with exponent $\beta=0.240\pm0.011$, or that it has the standard time dilation ($\alpha=1$) and  a large intrinsic dependence with exponent $\beta=1.240\pm0.011$. Since the initial division of the epoch differences by ${(1+z)}$ could produce the slope $\beta=1.0$ this analysis shows strong support for zero time dilation.
\par
Note that if there is no time dilation and the effects of auxiliary parameters are small, then the SALT2 stretch factors are estimates of the true width.
\section{\bf{THE ANALYSIS OF TYPE Ia SUPERNOVAE LIGHT CURVES}}
\label{s5}
\subsection{The raw observations}
\label{s5.2}
\citet{Crawford17} describes the selection and analysis of the original observations of type Ia supernovae light curves that have been selected by \citet{Betoule14}, who have provided an update of the \citet{Conley11} analysis with better optical calibrations and more supernovae. This JLA (Joint Light-curve Analysis) list sample has supernovae from the Supernova Legacy Survey (SNLS),nearby supernovae (LowZ), the Sloan Digital Sky Survey (SDSS) \citep{Holtzman08,Kessler09} and those observed by the  (HST) (Hubble Space Telescope) \citep{Riess07, Jones13}. Also included are 1169 supernova from the Pan-STARRS supernova survey \citep{Kaiser10, Jones18, Scolnic18}.  The sources of the raw observations are listed in the Appendix~\ref{a1}.
\par
For each type Ia supernova, the data used here was, for each filter, a set of epochs with calibrated flux densities and  uncertainties. The observations taken with the $U$ and $u$ filters are very noisy,  and following \citet{Conley11} and \citet{Betoule14}, the observations for these filters  were not used. Table~\ref{t1} shows the number of  type Ia supernovae, number of accepted supernovae, and number of accepted filter data for each survey. Most of the missing filter sets were because they did not have at least one observation  prior to five days before the peak flux density epoch and at least one  five days after  the peak.
\begin{table}
\caption{Surveys and numbers}
\centering
\label{t1}
\begin{tabular}{lrrr}
Survey & N$_{supernovae}$ & N$_{accepted}$ & N$_{filters}$ \\ \hline
 LOWZ &  241 &   96 &  290\\
 SDSS &  625 &  485 & 1563\\
 SNLS &  252 &  196 &  463\\
 PS1MD& 1169 &  731 & 1362\\
 HST  &   46 &    6 &    6\\
 Total& 2333 & 1514 & 3684\\
\hline
\end{tabular}
\end{table}
\subsection{Type Ia supernovae reference template}
\label{s5.1}
The essential aim of this analysis is to determine $\alpha$ and $\beta$ in Eq.~\ref{f2}, by examining the raw observations of type Ia supernovae. A critical part of any investigation into type Ia supernovae light curves is to have a reference template. In order to remove any possible bias,  a standard independent template, the $B$ band Parab-18 from Table~2 \citet{Goldhaber01} which has a half-peak width of 22.3 days has been used. Then the procedure is (for each supernova)  to determine the observed width of the light curve for each filter,  relative to the template light curve, and then  estimate $\alpha$ and $\beta$ from all the widths as a function of redshift.
\subsection{The analysis of raw observations}
\label{s5.3}
\begin{figure}
\includegraphics[width=\columnwidth]{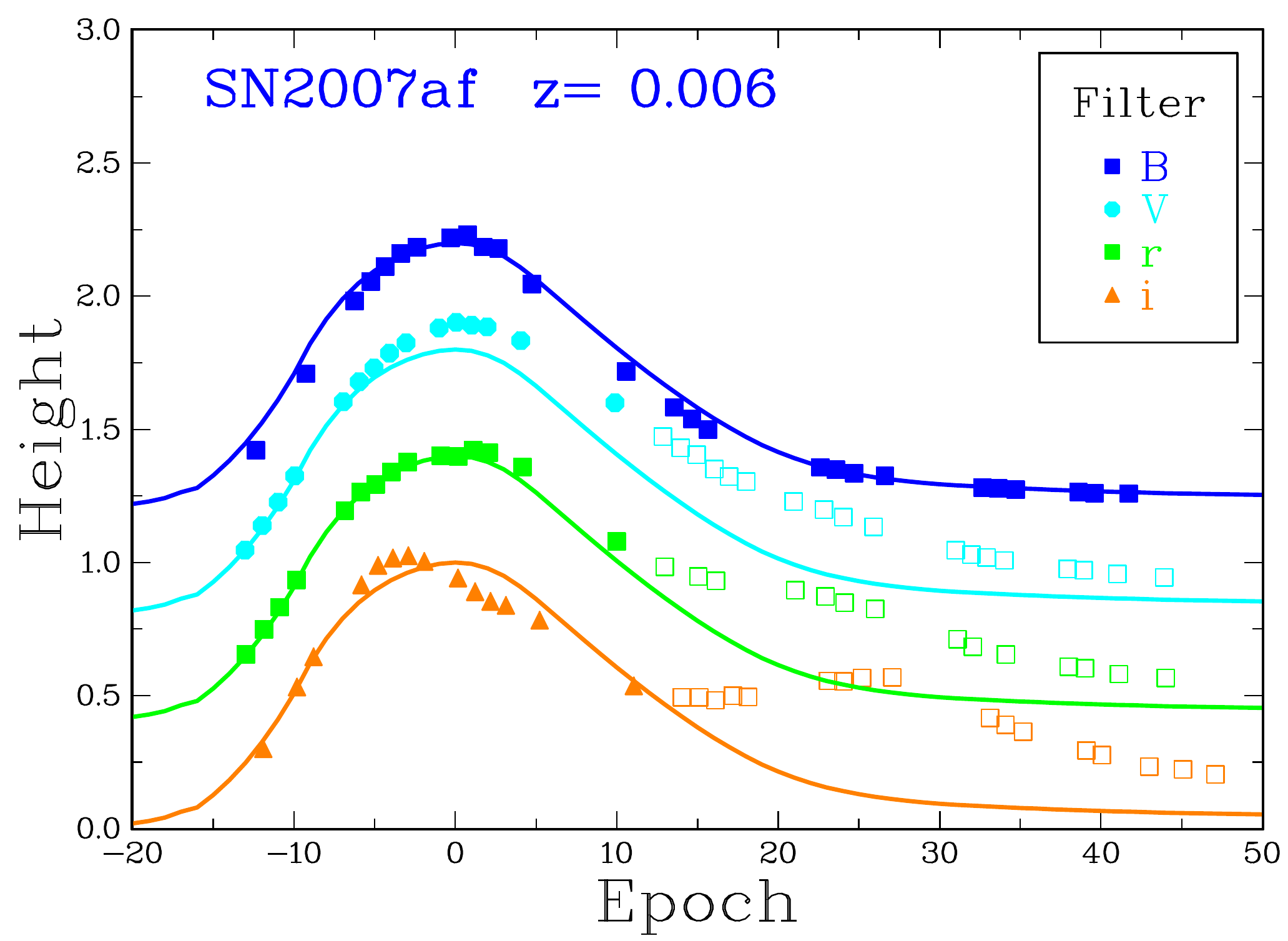}
\label{fig2}
\caption{The light curves for the SNLS type Ia supernova SN2007af. Valid points are shown as full squares and invalid points as open  squares. To avoid confusion, the  filter results have been vertically displaced, The secondary peak is clearly apparent.}
\end{figure}

As an example of a typical type Ia supernova Figure~\ref{fig2} shows the light curves for four filters for the SNLS supernova SN2007af with the filters used being shown in the legend \citep{Goobar11}. Accepted data points are shown as coloured symbols whereas  the rejected points are shown with an open square.
\par
The first feature to notice is that the epoch of the peak flux density depends on the filter type and is therefore a function of the intrinsic wavelength. Secondly, there is a secondary peak at about 25 days after the first peak for the longer wavelength filters. Although this second peak is intrinsic to the supernova, it does not appear to be very consistent \citep{Elias81,Meikle00,Goobar11}. Consequently, as shown in Figure~\ref{fig2}, all filters, except $B$,  $j$, and  HST, had their epochs more than 15 days after the main peak rejected.
\par
Unfortunately, the direct analysis of the data to obtain the epoch of the peak flux density, the value of the peak flux density and the light curve width using a  $\chi^2$ method has an intrinsic problem in that position of the peak flux density and the width are not completely  independent. However the value of the peak flux density is almost independent of the width estimate.
\par
The first step was to estimate the value of the peak flux density using a minimum $\chi^2$ procedure. Next the program uses  the reference light curve and the ratio of the flux density to the peak flux density to obtain a flux density epoch. This epoch has an uncertainty equal to the flux density uncertainty divided by the absolute value of the slope of the reference light curve at that epoch. Then a simple weighted regression of the observed epochs verses the flux density epochs provides the peak flux density offset and the width of the light curve. This estimate of the peak flux density offset was ignored. However it was found by minimising the standard $\chi^2$ for the flux densities that was calculated using  the estimates for the widths and the flux densities.
\par
This method has  the  bonus of providing uncertainty estimates for the widths. Note that  each  supernova had separate values for  the peak flux density and the width for each filter.
\par
One problem that was noticed is that the range of the uncertainties in the flux densities for many filters was too large. There were 336 cases out of 42,818 filter sets where the ratio of the smallest uncertainty to the largest uncertainty was less than 10\%. The problem is that one of these very precise  flux densities could have a weight one hundred times larger than other flux densities which could produce anomalous results. The observation method for most of these supernovae is to observe the same patch of sky with the same telescope and settings for each epoch. Although there can be nights with bad seeing, the expected uncertainty in the flux density is the same for each observation. Consequently all the flux density uncertainties were replaced by a common value of 3\% of the peak flux density.
\par
After all the parameters were estimated for a particular filter and the supernova, the flux density for each epoch was tested to see if it was an outlier. This was done by computing a value
\begin{equation}
l_i = f_i/f_{peak} - h_i,
\end{equation}
where for each epoch,  $f_i$ is its flux density, $f_{peak}$ is the peak flux density,  $h_i$ is the height of the reference light curve at that epoch, and $i$ is its index. Then an epoch it was rejected as an outlier if the value $|l_i-\bar{l} |$,  where $\bar{l}$ is the average for all the other epochs $l_i$, was greater than five times the rms for all the other epochs. The epoch with the largest discrepancy was eliminated, then a full analysis was repeated and this continued until there were no more outliers.  Out of 30,850 accepted epochs, there were 964 (3\%) outliers.
\par
The major selection criteria for each valid supernova was that, for each filter, there was at least one epoch less than five days from the peak epoch and one that was five days greater than the peak epoch, and  that there were at least 4 valid epochs. And in order to show a reasonable fit to the light curve the width uncertainty must be greater than 0.005 and less than 0.3.
\subsection{Light curve widths}
\label{s5.4}
\begin{figure}
\includegraphics[width=\columnwidth]{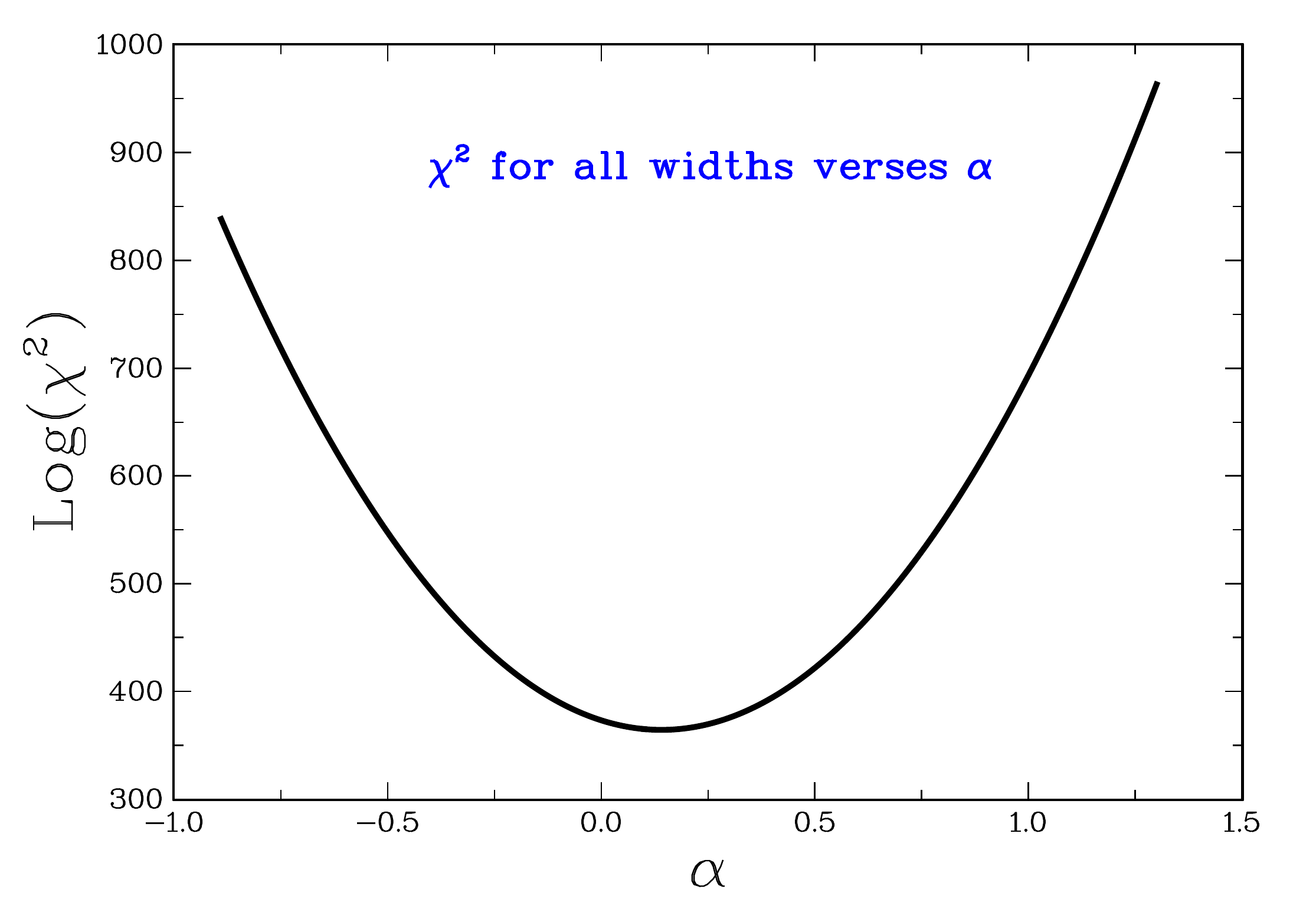}
\label{fig3}
\caption{The $\chi^2$ fit for light curve widths  as a function of $\alpha$. For each $\alpha$ the intrinsic width is set to be $\beta=\alpha - 0.352$.}
\end{figure}\
\par
\begin{figure*}
\plottwo{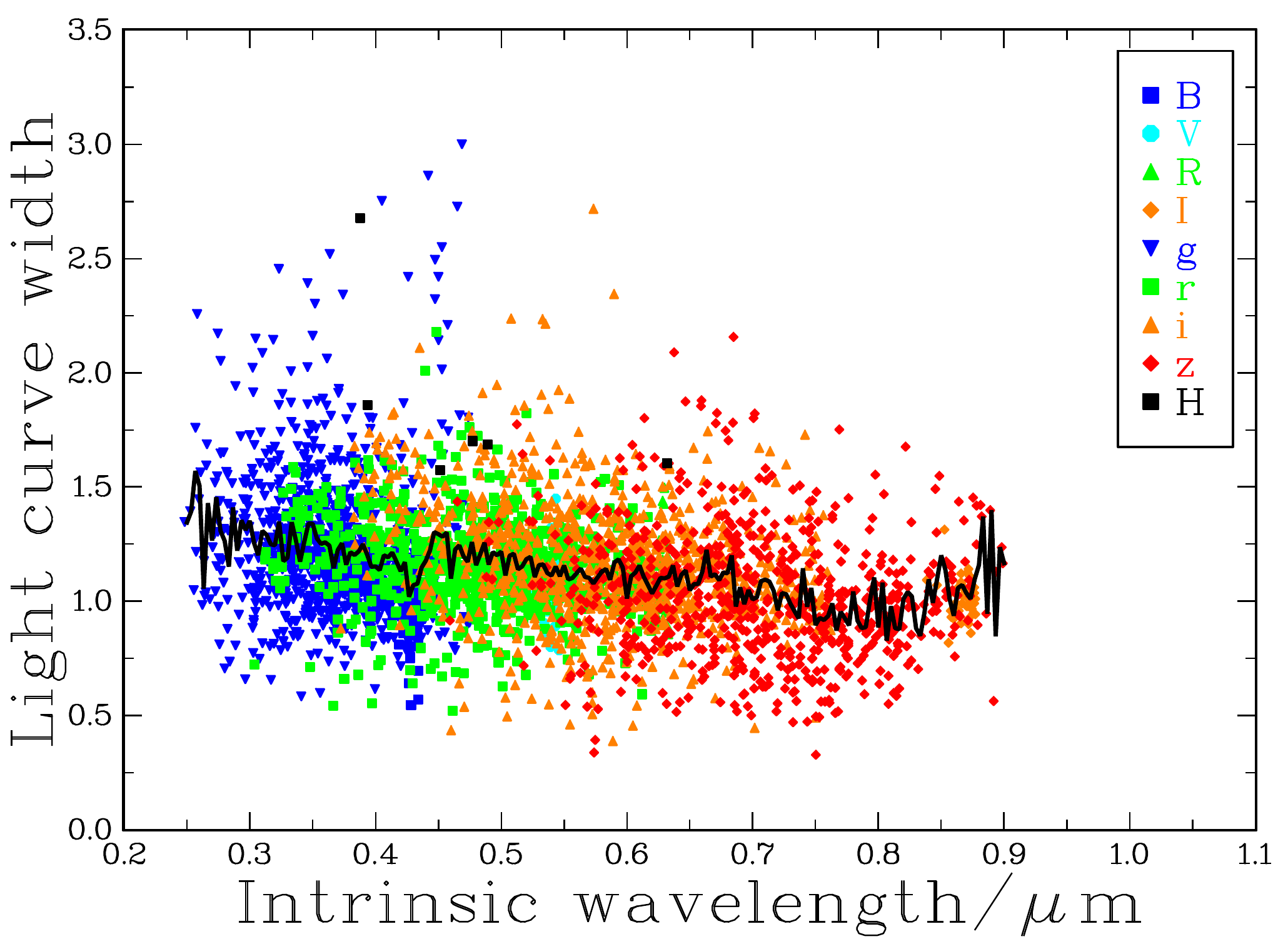}{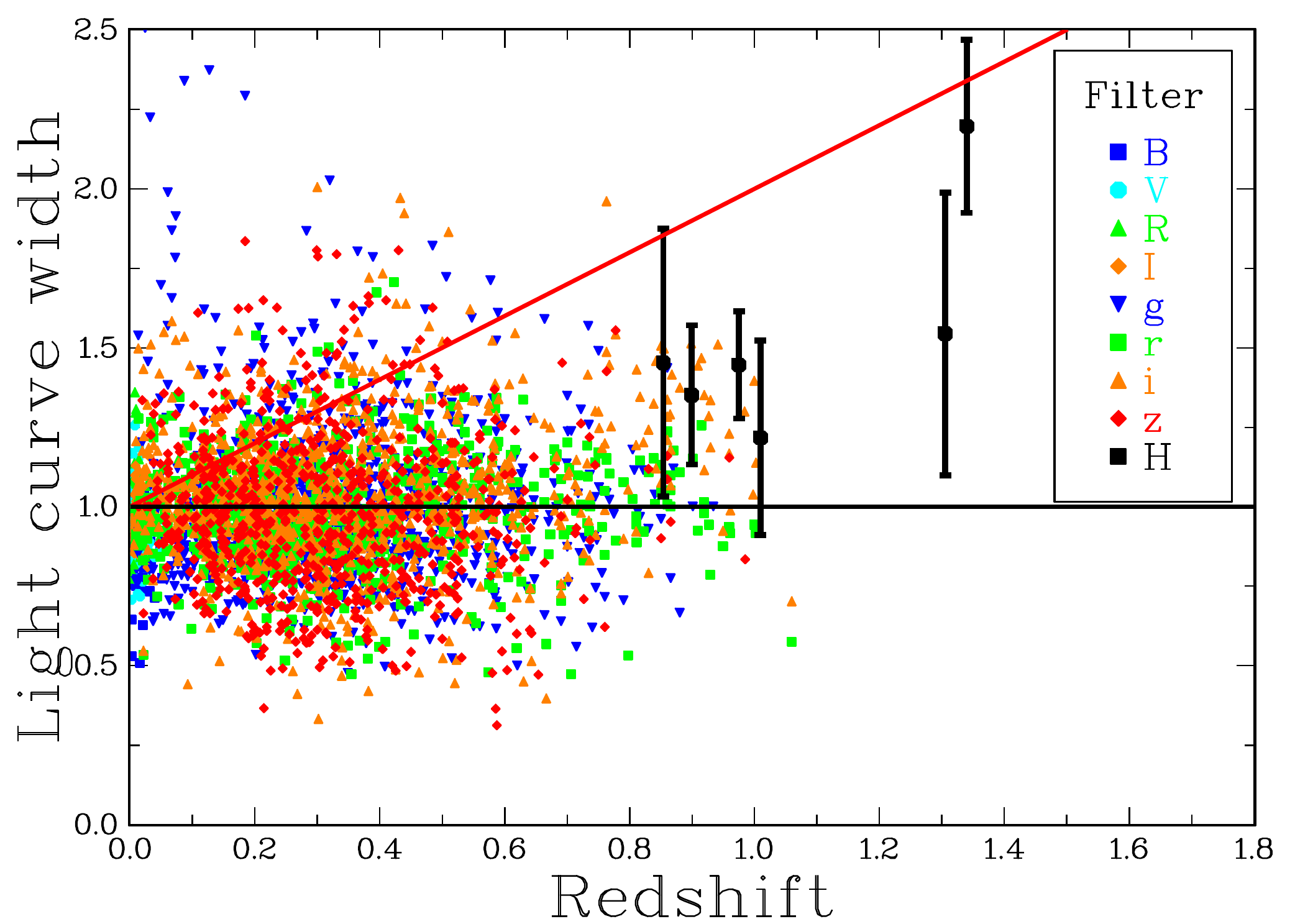}
\caption{ This is a plot of the observed  widths of  type Ia the light curves  as a function of intrinsic wavelength and redshift.   The symbol and colour for each filter are shown in the legend.   The black line is the average value as a function of $\epsilon$. For the right figure, the black line shows the width for $\alpha=0$ and the red line shows the width for standard time dilation $\alpha=1$. The symbol and colour for each filter are shown in the legend. In order to avoid clutter only the HST observations (shown in black)  have  uncertainty estimates.\label{fig4}}
\end{figure*}
Because of Eq.~\ref{f3} there is confusion in measuring $\alpha$ and $\beta$ from the raw observations. However Eq.~\ref{f3} shows that a regression
 of the logarithm of the width verses the logarithm of $(1+z)$ will provide an estimate of the combined $\alpha-\beta$. The result for this raw width is
\begin{equation}
\label{e6}
\alpha-\beta= 0.324\pm0.025.
\end{equation}
One way to determine which time dilation is appropriate for these observations is to plot the $\chi^2$ of the fit as a function of $\alpha$ where for each $\alpha$ we put $\beta=0.324-\alpha$, thus preserving the raw width. The  $\chi^2$ function is
\begin{equation}
\label{e7}
\chi^2= \sum_i[\log(w_i) - A - \alpha\log(1+z_i) - \beta\log(\lambda/(1+z_i))]^2,
\end{equation}
where the summation is over all the observations with index $i$, $w_i$ is the width and $A$ is a dummy variable that is determined for each $\alpha$. The result is shown in Figure~\ref{fig3}. Clearly the best fit is when $\alpha\approx 0.2$ which is compatible with  zero time dilation which implies that $\beta\approx -0.124$.
\par
This is supported by another estimate of $\beta$ from the average value of the logarithm of the ratio of the widths of two filters from the same supernovae. Although this method assumes that both filters have the same width it is independent of the value of the width.  For 496 supernovae that had valid observations in the $g$ and $z$ filters the estimate is $\beta=-0.302\pm0.023$.
\par
A more general and better analysis is to assume that $\alpha=0$ and the get the intrinsic widths a  function of $\epsilon$ rather than restrict them to a simple function $(1+z)^\beta$. This is easily done by averaging the widths for each value of  $\epsilon$. Figure~\ref{fig4} shows a scatter distribution of intrinsic width for each observation together with its average value. For reference purposes the value of the average exponent $\beta$ was also estimated. For 3,684 accepted widths the result was $\beta=0.212\pm0.007$. This value is in excellent agreement with the SALT2 value of $\beta=0.240\pm0.011$(Section~\ref{s4.2}).
\par
The final step is the estimate $\alpha$ by a regression of the logarithm  of the raw widths corrected for intrinsic width verses $log(1+z)$. The correction was done by dividing the observed width by the intrinsic width. The regression equation is
\begin{equation}
\label{f5}
w_{cosmological}=(-0.044\pm0.004)(1+z)^{0.083\pm0.024}.
\end{equation}
This width is self-consistent with zero  which strongly favours a static universe with no time dilation. Figure~\ref{fig4} shows the intrinsic distribution and the  plot of the cosmological widths for 3,684 filters from 2,333 type Ia supernovae.
\subsection{Redshift dependence of the  light curve widths}
\label{s5.5}
The observed light curve for each of four redshift ranges was computed for each epoch in the range from -15 days to 40 days from the peak flux-density epoch. This was done by selecting all relative flux densities  that were within one day of this epoch and setting the value of the light curve to be the median of these selected flux densities. The median was used because it is insensitive to extreme values.
Although this method is using the same data, its advantage is that it depends only on the relative flux density for each epoch and does not depend on the fitting procedure. This is similar to the type of analysis done by \citet{Goldhaber01} and \citet{Blondin08}.
\begin{figure}
\includegraphics[width=\columnwidth]{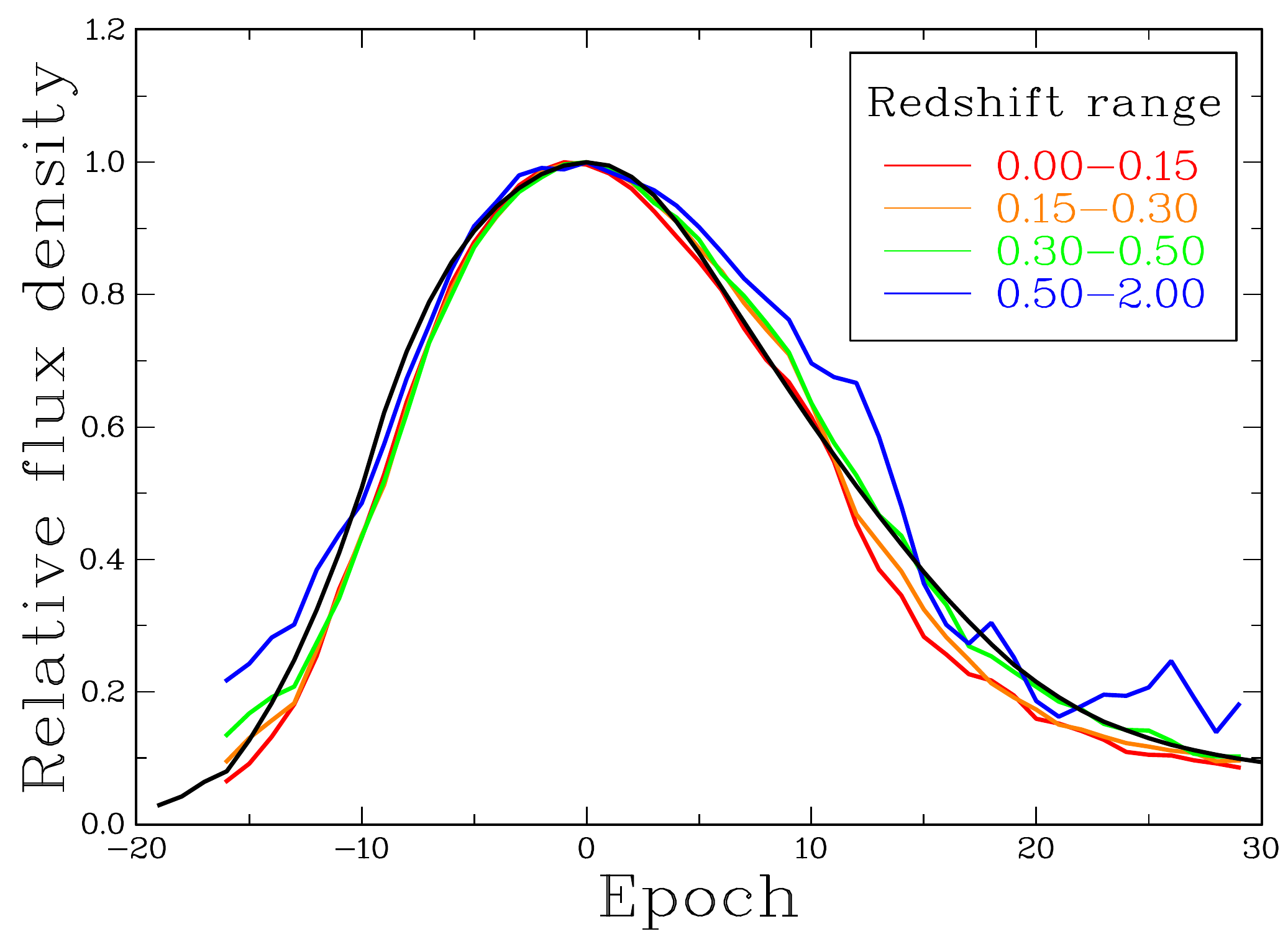}
\caption{\label{fig6} The type Ia supernovae light curves for four redshift ranges for the static model. The legend shows the colour for each redshift range. The template light curve is shown in black. The epochs have been corrected for the intrinsic width by multiplying each epoch difference by  the appropriate intrinsic width}
\end{figure}
\par
The results are presented graphically in Figure~\ref{fig6} which shows the average light curve for four ranges of redshift. The black  curve shows the master template light curve. Table~\ref{t2} shows the redshift range, the mean redshift, the number of points, and the average width for each range. Note that the observed width for each epoch has been corrected for its intrinsic component by multiplying the epoch distance from the peak flux density epoch by the intrinsic width show by the black line in Figure~\ref{fig4}.
\begin{table}
\caption{Light curve widths for four redshift ranges}
\centering
\label{t2}
\begin{tabular}{cccc}
\hline
Range & $\bar{z}$ & number & Width \\ \hline
 0.00-0.15& 0.069& 25 & $0.903\pm0.005$\\
 0.15-0.30& 0.225& 25 & $0.954\pm0.009$\\
 0.30-0.50& 0.383& 25 & $1.059\pm0.024$\\
 0.50-1.30& 0.649& 25 & $1.195\pm0.033$\\
\hline
\end{tabular}
\end{table}
The power law fit for these four widths with respect to $(1+z)$ has an exponent of $0.023\pm0.012$ which has  a negligible  dependence on redshift.
\subsection{Type Ia supernovae peak magnitudes}
\label{s5.7}
\begin{figure*}
\plottwo{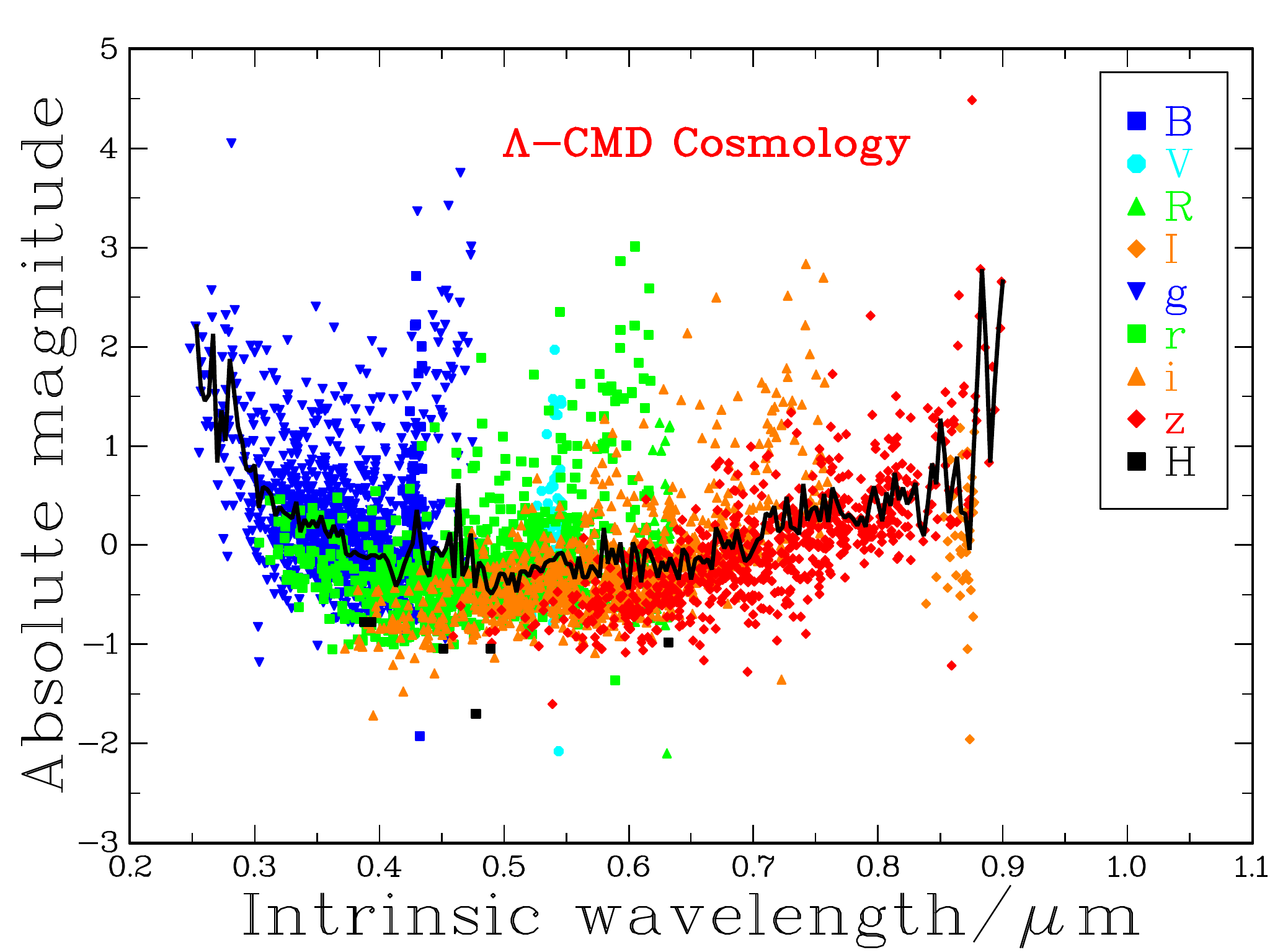}{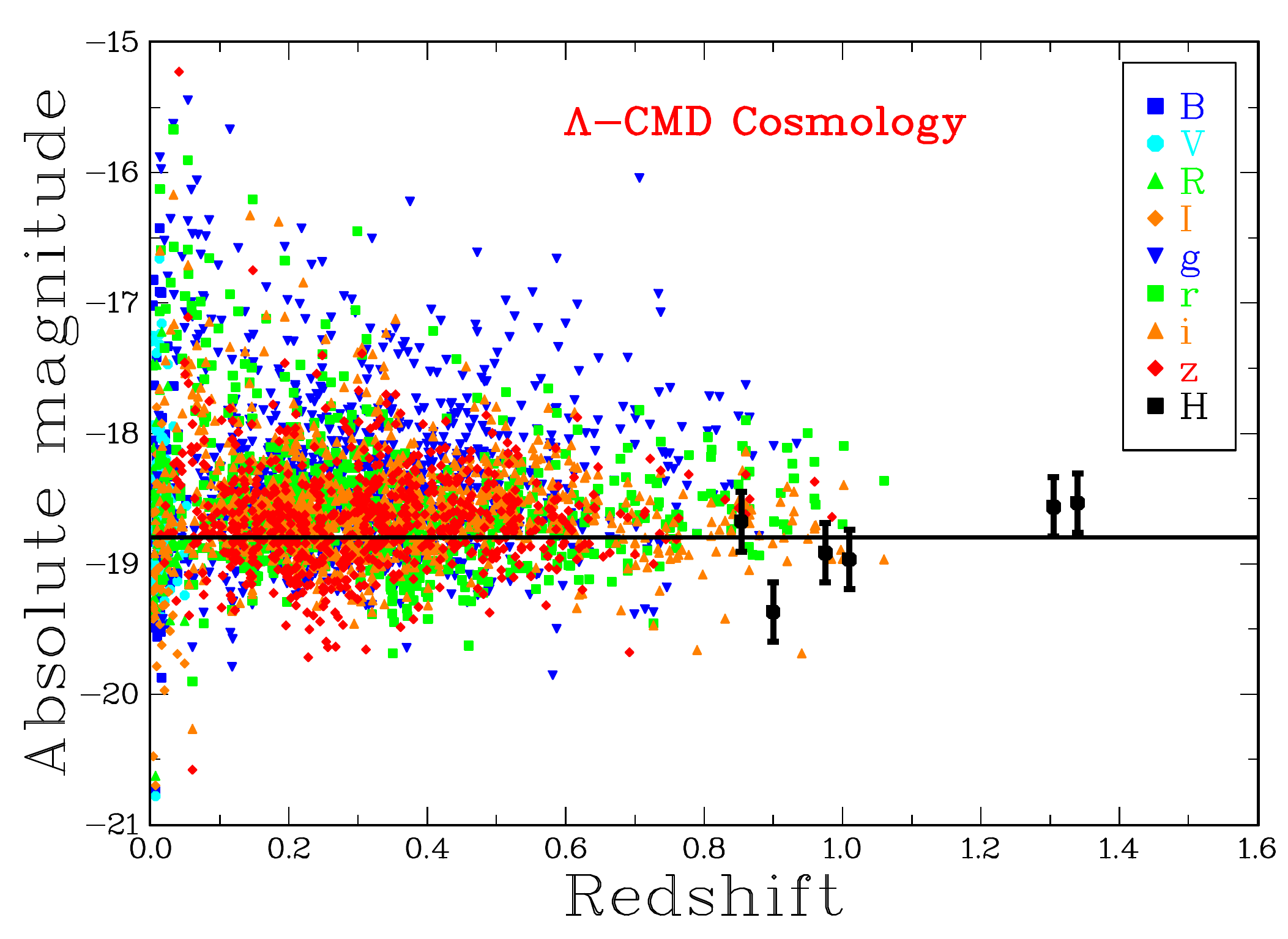}
\caption{\label{fig7} Scatter plots of the intrinsic absolute magnitude as a function of $\epsilon$ and the cosmological peak absolute magnitude as a function of redshift for the $\Lambda$-CDM  model. The black line in the left figure shows  the average value of the intrinsic magnitude.}
\end{figure*}
\begin{figure*}
\plottwo{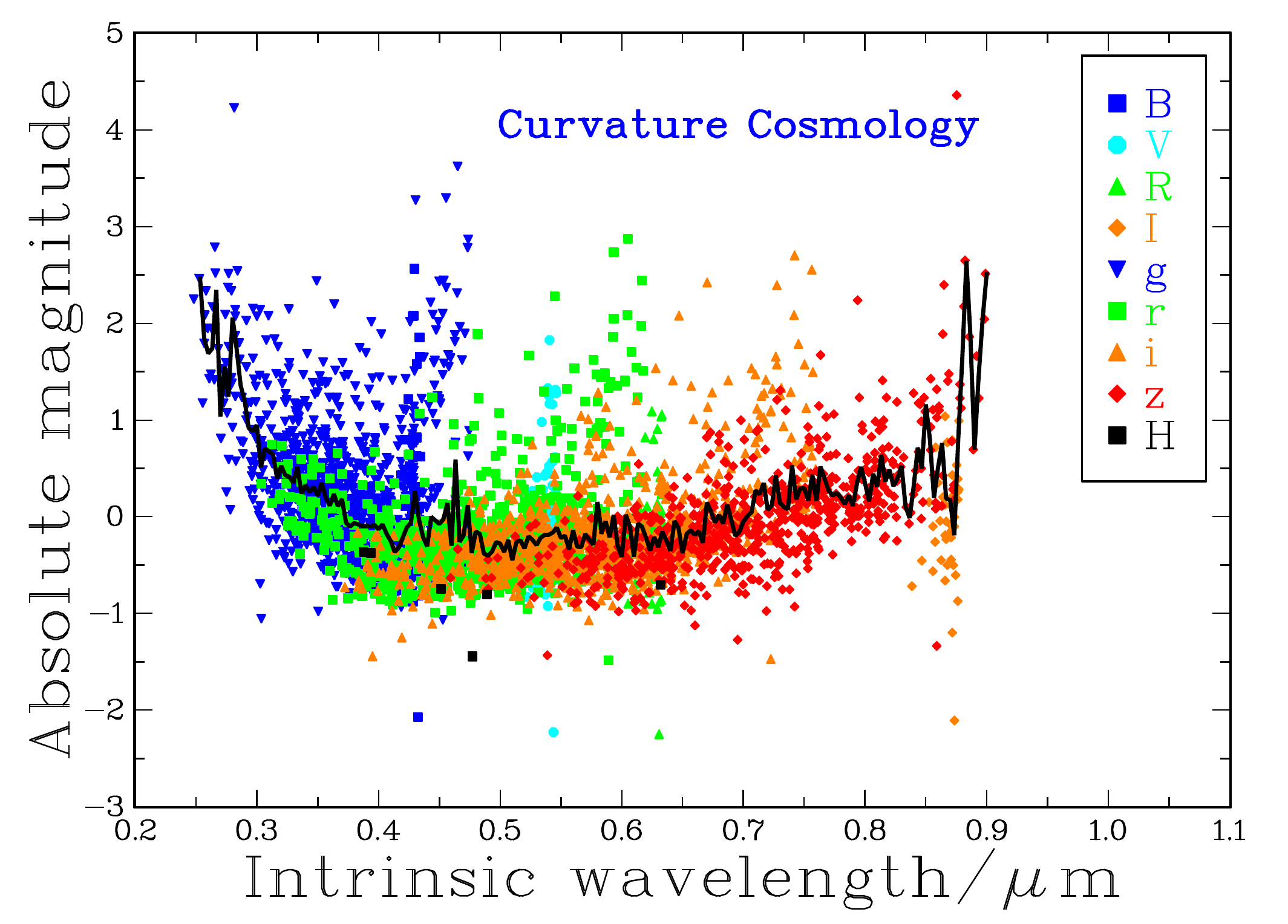}{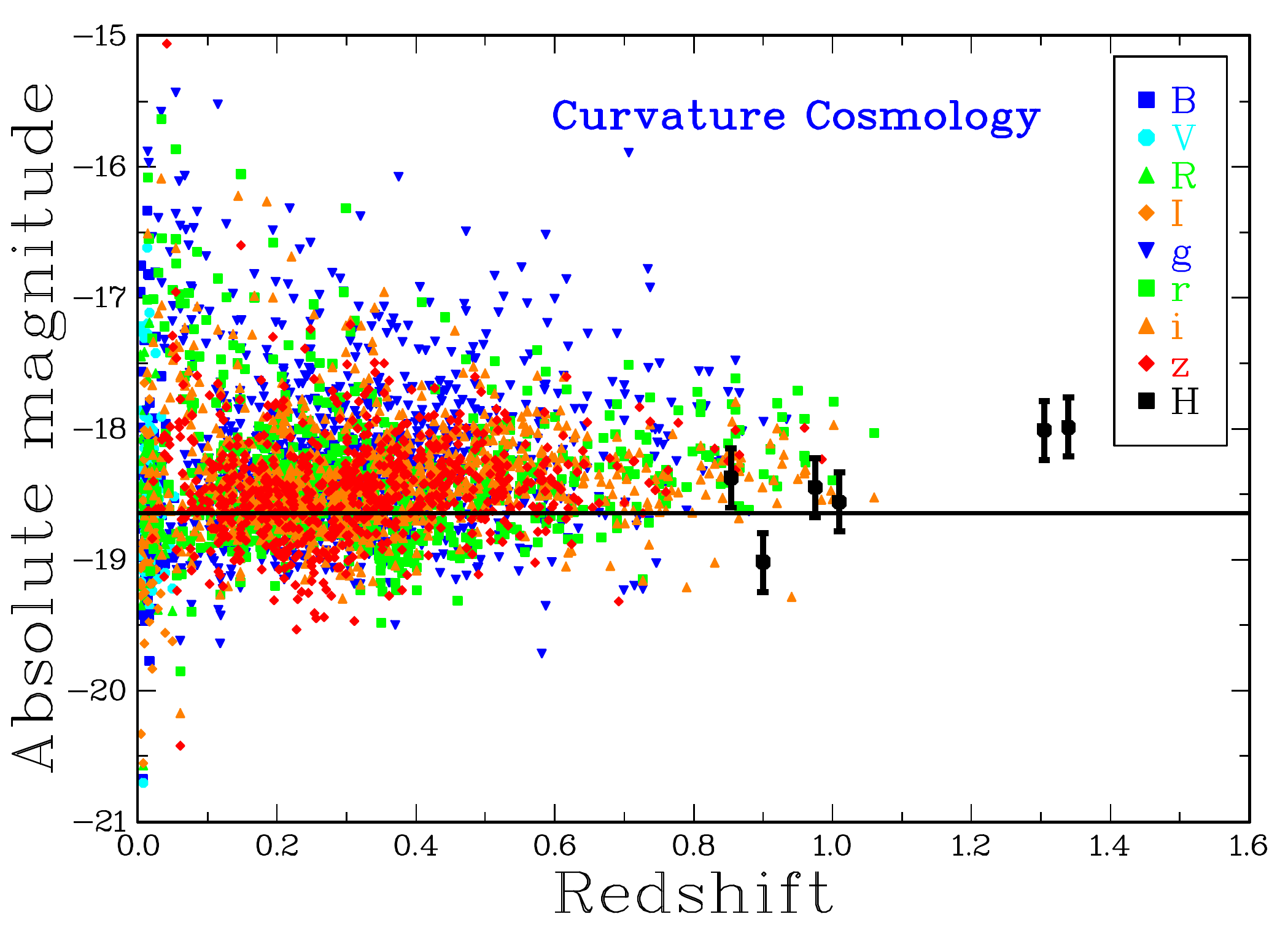}
\caption{\label{fig9} Scatter plots of the intrinsic absolute magnitude as a function of $\epsilon$ and the cosmological peak absolute magnitude as a function of redshift  for Curvature Cosmology. The black line in the left figure shows  the average value of the intrinsic magnitude.}
\end{figure*}
Since the observed light curve is  the intrinsic light curve multiplied by time dilation, then the observed flux density  is the product of the intrinsic flux density and the cosmological scaling factor. There is strong evidence from nearby supernovae that the expected peak absolute magnitude  is the same for each supernova. Clearly, this commonality requires a valid distance modulus coming from a valid cosmological model.  The analysis is done for the standard $\Lambda$-CDM  cosmology and for a static cosmology. A suitable static model is Curvature Cosmology which is a complete cosmology that shows excellent agreement with all major cosmological observations. A brief description of the cosmology  is given later (Section~\ref{s8}).
\par
For convenience all the flux densities were converted into AB magnitudes. The analysis  procedure used is identical to that used above for the light curve widths which starts with the measurement of the intrinsic absolute magnitudes  which are then subtracted from the raw magnitudes to get the cosmological magnitudes.  Table~\ref{t3} shows the results for both cosmologies for the exponent $\alpha$ from the regression of the absolute peak magnitudes verses $-2.5\log_{10}(1+z)$. The top row is for the raw peak magnitudes and the bottom row is for the cosmological peak magnitudes.
\begin{table}
\centering
\caption{Peak magnitude exponents}
\label{t3}

\par\begin{tabular}{lrr}
\hline
 Parameter               & $\Lambda$-CDM    & Curvature Cosmology \\ \hline
 $\alpha_{raw}$          & $ 0.182\pm0.030$ & $-0.385\pm0.043$ \\
 $\alpha_{cosmological}$ & $-0.014\pm0.030$ & $0.050\pm0.030$ \\
\hline
\end{tabular}
\end{table}
Both models are consistent with the peak absolute magnitude being independent of redshift. Since the type Ia supernovae observations have been a major contribution the   $\Lambda$-CDM model it is not surprising that it has a good fit to this data.
\par
The good fit of  Curvature Cosmology is a strong endorsement of this model. Note that it was formulated long before good supernovae observations became available and its distance modulus has no fitted parameters except for $H$ (Eq.~\ref{e10}) which an additive constant.  Figure~\ref{fig7} shows the intrinsic absolute magnitude as a function of $\epsilon$ and the cosmological peak absolute magnitude as a function of redshift for the $\Lambda$-CDM  model. Figure~\ref{fig9} shows the same results for Curvature Cosmology.
\section{\bf{GAMMA RAY BURSTS}}
\label{s6}
The website of the {\it Neil Gehrels} Swift Observatory, which runs the Swift satellite, that contains the Burst Alert Telescope (BAT) describes them as: ``Gamma-ray bursts (GRBs) are the most powerful explosions the Universe has seen since the Big Bang. They occur approximately once per day and are brief, but intense, flashes of gamma radiation. They come from all different directions of the sky and last from a few milliseconds to a few hundred seconds.'' An important characteristic of  the BAT is that it has a  photon counting detector  \citep{Barthelmy05} that detects photons in the 15-150 keV energy range with a resolution of about 7 keV. It can also image up to 350 keV without position information. An important parameter for each burst is T$_{90}$ which is a measure of the burst duration. The start and end times of T$_{90}$ are defined as the times  the fraction of photons in the accumulated light-curve reaches 5\% and 95\%.
\par
The Third Swift Burst Alert Telescope Gamma-Ray Burst Catalog \citep{Lien16} states that ``Many studies have shown that the observed burst durations do not present a clear-cut effect of time dilation for GRBs at higher redshift." Indeed the upper panel of their Figure~25 shows that there is no obvious trend of the burst length with redshift except for a decrease in the number of short bursts with larger redshifts. This shows some support for the ``tip-of-the-iceberg" effect which is sometimes used to explain the lack of strong time dilation in the GRB durations. However, there is no obvious change in the duration of longer bursts with redshift.
\par
Now the number  distribution of photons in GRB bursts is close to a power law with an exponent of about -1.6. As the redshift increases, the number of detectable photons will rapidly decrease as many more photons will be below the detector limit. If we assume that the distribution of photons as a function of energy is independent of the position of the photon in the burst, then there should be no expected change in T$_{90}$ with redshift. On the other hand, if the higher energy photons are clustered towards the centre of the GRB, then the intrinsic T$_{90}$ should decrease with increasing redshift. Consequently,  we would expect to see the normal time dilation or maybe a little less in the T$_{90}$ measurements.
\par
This analysis directly examines  the  exponent of a power-law regression of measured T$_{90}$ of raw GRB data (c.f. Appendix~\ref{a2}), that had burst durations above 2 seconds, as a function of $(1+z)$\citep{Lien16}.  Since there were no  T$_{90}$ uncertainties provided, the analysis used an unweighted regression.
The power-law fits were done for the T$_{90}$ duration with the exponent shown in row~1 of Table~\ref{t5} which is consistent with no time dilation. The problem with this and similar analyses is that the variables have a very large scatter in values which would require very large numbers of GRB to achieve absolutely conclusive results.
\par
In a recent analysis \citet{Zhang13} claim that the GRB T$_{90}$ widths are consistent with an expanding universe.  They measured T$_{90}$ in the observed energy range  between $140/(1+z)\,$keV and $350/(1+z)\,$keV, corresponding to an  intrinsic energy range of $140-350\,$keV. Their exponent for these selected is $0.94\pm0.26$ which is consistent with the standard expanding model.
\par
\begin{figure}
\includegraphics[width=\columnwidth]{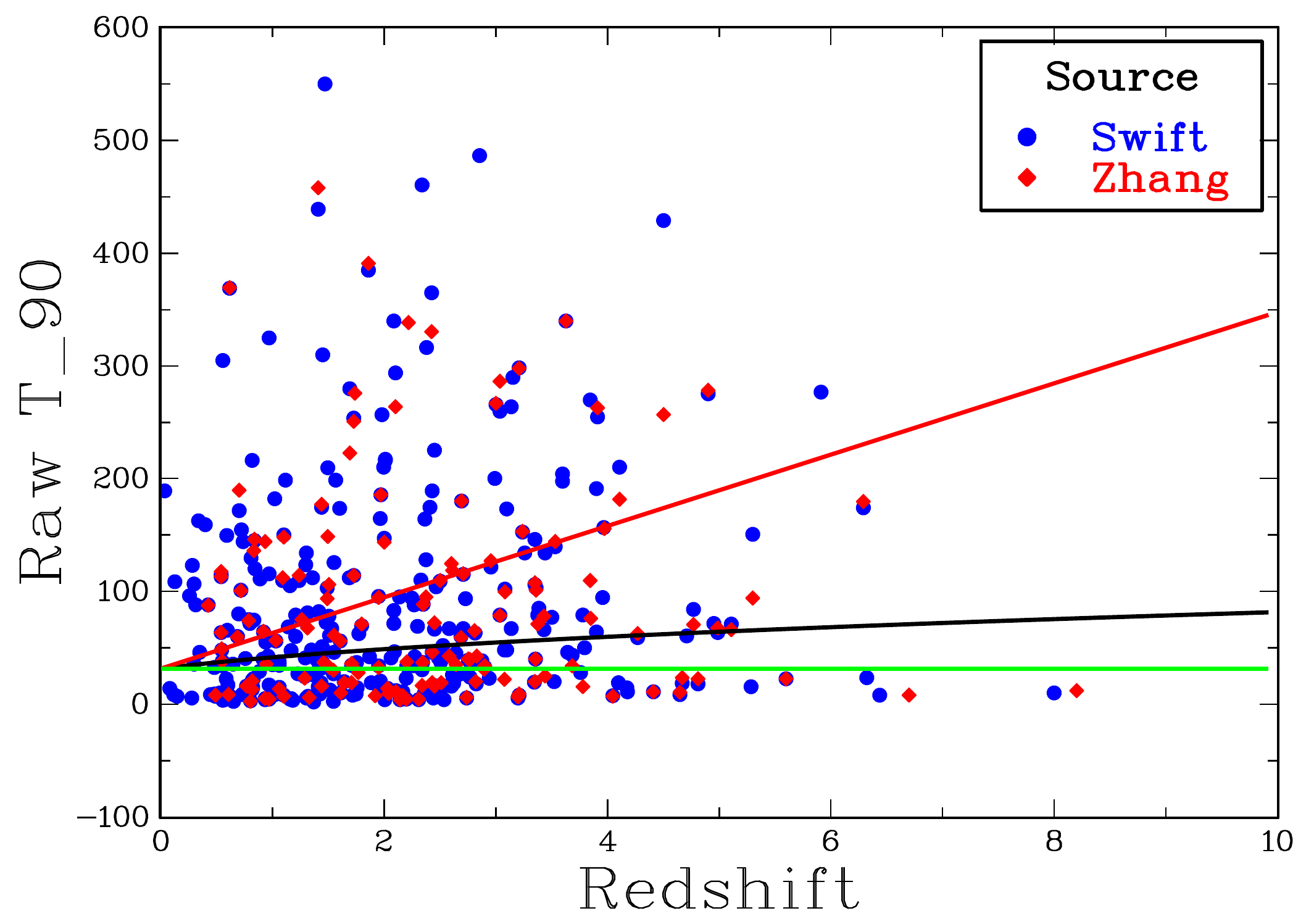}
\caption{\label{fig20}A plot of T$_{90}$  as a function of redshift. The green line shows the line for no time dilation,  the red line shows the line for standard time dilation,  and the black line shows a time dilation with the fitted  exponent of $0.39$. The Swift data are shown in blue filled circles and the Zhang \citep{Zhang13} data are shown as red diamonds.}
\end{figure}

My reanalysis of their raw, T$_{90}$ widths using the data in their Table~1, is shown in columns~2 and 3 of Table~\ref{t5} and are consistent with no time dilation. The Swift and \citet{Zhang13} unselected  T$_{90}$ widths are displayed in Figure~\ref{fig20}. Although they have many common GRB, there are small differences in the T$_{90}$ widths. This is because \citet{Zhang13} have used their own analysis of the original data to get their own values for T$_{90}$ widths. Examination of  Figure~\ref{fig20} shows the large scatter of the T$_{90}$ widths and  it also shows that they are consistent with no time dilation and are unlikely to be consistent with standard time dilation.
\par
My determination of the exponents of their energy selected widths as a function of $(1+z)$ is shown in rows~4 and 5 of Table~\ref{t5}. The unweighted result in row~4  agrees with their result. However the  exponent for the weighted analysis shown in row~5 is consistent with no time dilation.
\begin{table}
\centering
\caption{Exponents for redshift dependence of GRB}
\label{t5}
\begin{tabular}{clccr}
\hline
\hline
Row & Data        & Weight$^{a}$ & N   &  Exponent\\ \hline
1   & Swift       & U            & 298 & $ 0.39\pm 0.17$ \\
2   & Zhang$^{b}$ & U            & 139 & $ 0.10\pm 0.26$ \\
3   & Zhang$^{b}$ & W            & 139 & $-0.16\pm 0.20$ \\
4   & Zhang$^{c}$ & U            & 139 & $ 0.94\pm 0.26$ \\
5   & Zhang$^{c}$ & W            & 139 & $ 0.31\pm 0.23$ \\
\hline
\end{tabular}
\begin{flushleft}
$^{a}$ ~U denotes an unweighted fit, W denotes weighted\\
$^{b}$ ~Raw T$_{90}$ from \citet{Zhang13} in their Table~1\\
$^{c}$ ~T$_{90}$,$_z$ for intrinsic energy range of 140-350 keV.
\end{flushleft}
\end{table}
\par
The use of  energy selection for T$_{90}$ implies that there is an  intrinsic  dependence of burst duration on the photon energy. Since the BAT has a photon counting detector, any measurement of T$_{90}$ is independent of the selected photon energies. The only restriction is that the photon energies must be within the detector limits.  Thus BAT does not have the energy selection that is necessary for this analysis. Furthermore, it is difficult to understand how any subset of photons that are detected  can have a different time dilation from the rest of the photons in the same GRB. If we ignore the energy-selected \citet{Zhang13} results, the conclusion is that the burst length of GRB is consistent with no time dilation and has very little support for the standard model.
\section{\bf{CURVATURE COSMOLOGY}}
\label{s8}
Curvature Cosmology \citep{Crawford87a,Crawford87b,Crawford91,Crawford93,Crawford95a,Crawford95b,
Crawford99,Crawford06,Crawford09a,Crawford09b,Crawford10} is a complete cosmology for a static universe that shows excellent agreement with all major cosmological observations without needing dark matter or dark energy.(Note that \citep{Crawford10} is an update with corrections of the previous work.) This cosmology depends on the hypotheses of Curvature Redshift and Curvature Pressure described below.
\par
The basic cosmological model is one in which the cosmic plasma dominates the mass distribution and hence the curvature of space-time. In this first-order model, the effects of galaxies and stars are neglected. The geometry of this cosmology is that of a three-dimensional surface of a four-dimensional hyper-sphere. It is almost identical to that for Einstein's static universe. For a static universe, there is no ambiguity in the definition of distances and times. One can use a cosmic time and define distances in light travel times or any other convenient measure.  Curvature Cosmology obeys the perfect cosmological principle of being statistically the same at all places and at all times.
\subsection{Curvature Redshift}
\label{s8.1}
The derivation of Curvature Redshift is based on the fundamental hypothesis of Einstein's general theory of relativity that space-time is curved. As a consequence, for positive curvature, the trajectories of initially parallel point particles, geodesics, will move closer to each other as time increases. Consequently  the cross-sectional area of a bundle of geodesics  will slowly decrease.
\par
In applying this idea to photons, we assume that a photon is described in quantum mechanics as a localised wave where the geodesics correspond to the rays of the wave. Note that this wave is quite separate from an electromagnetic wave that corresponds to the effects of many photons. It is fundamental to the hypothesis that we can consider the motion in space-time of individual photons. Because the curvature of space-time causes the focussing of a bundle of geodesics, this focussing also applies to a wave. As the photon progresses, the cross-sectional area of the wave associated with it will decrease. However, in quantum mechanics properties such as angular momentum are computed by an integration of a radial coordinate over the volume of the wave. If the cross-sectional area of the wave decreases, then the angular momentum will also decrease. However, angular momentum is a quantised parameter that has a fixed value.
\par
The solution to this dilemma is that the photon  splits into two very low-energy photons  and a third that has the same direction as the original photon and nearly all the energy. It is convenient to consider the interaction as a primary photon losing a small amount of energy to two secondary photons. This energy loss will be perceived as a small decrease in frequency. By symmetry the two secondary photons with identical energies are emitted at right angles to the trajectory, which means that there is no apparent angular scattering.
\par
Since in quantum mechanics electrons and other particles are considered as waves, a similar process will also apply. It is argued that electrons will interact with curved space-time to lose energy by the emission of very low-energy photons.
\par
From \cite{Crawford10} we get the basic equation for the fractional change in energy of the photon. This is based on the equation of geodesic deviation \citep{Misner73}.
\begin{equation}
\label{c1}
\frac{1}{E}\frac{dE}{ds}=-\left(\frac{8\pi G\rho}{c^2} \right)^{1/2}
=-1.366\times 10^{-13}\sqrt{\rho}\,\mbox{m}^{-1}.
\end{equation}
For many astrophysical types of plasma, it is useful to measure density by the equivalent number of hydrogen atoms per cubic metre: that is we can put  $\rho=$N\,M$_H$ and get
\begin{equation}
\label{c2}
\frac{1}{E}\frac{dE}{ds} =-\sqrt{\left(\frac{8{\pi}GNM_H}{c^2}\right)}=-5.59\times10^{-27}\sqrt{N}\, \mbox{m}^{-1}.
\end{equation}
The rate of energy loss per distance travelled depends only on the square root of the density of the material, which may consist of gas, plasma, or dust.
For many astrophysical plasmas the frequency of the emitted photons will be less than the plasma frequency and they will be absorbed and heat the plasma.
\par
Another important factor is that if there is any other competing interaction which occurs before the secondary photons are produced it will inhibit the Curvature Redshift. Such an interaction is the coherent multiple scattering that produces refractive index. This can be important for ground bases experiments and for radio frequency observations in the Galaxy. For example, most lower frequency radio observations in our galaxy will be unaffected by Curvature Redshift.
\subsection{Curvature Pressure}
\label{s8.2}
The hypothesis of Curvature Pressure is that for moving particles there is a pressure generated that acts back on the matter that causes the curved space-time. In this case, Curvature Pressure acts on the matter  that is producing curved space-time in such a way as to try to decrease the curvature. In other words, the plasma produces curved space-time through its density entering the stress-energy tensor in Einstein's field equations and the actions of the velocities of the plasma particles is to try and decrease this curvature..
\par
A simple cosmological model using Newtonian physics  illustrates some of the basic physics subsequently used to derive the features of Curvature Pressure. The model assumes that the universe is composed of gas confined to the three-dimensional surface of a four-dimensional hyper-sphere. Since the visualisation of four dimensions is difficult let us suppress one of the normal dimensions and consider the gas to occupy the two-dimensional surface of a normal sphere. From Gauss's law (i.e. the gravitational effect of a spherical distribution of particles with radial symmetry is identical to that of a point mass equal in value to the total mass situated at the centre of symmetry) the gravitational acceleration at the radius $r$ of the surface is normal to the surface, directed inward and it has the magnitude $\ddot r = -GM/r^2$ where $M$ is the total mass of the particles and the dots denote a time derivative. For equilibrium, and assuming all the particles have the same mass and velocity we can equate the radial acceleration to the gravitational acceleration and get the simple equation from celestial mechanics  of
\[
\label{c3}
v^2  = \frac{{GM}}{{r}}.
\]
The effect of this balancing of the accelerations against the gravitational potential is seen within the shell as a Curvature Pressure that is a direct consequence of the geometric constraint of confining the particles to a shell. If the radius $r$ decreases then there is an increase in this Curvature Pressure that attempts to increase the surface area by increasing the radius. For a small change in radius in a quasi-equilibrium process where the particle velocities do not change the work done by this Curvature Pressure (two dimensions) with an incremental increase of area $dA$ is $p_{\rm c}dA$ and this must equal the gravitational force times the change in distance to give
\[
p_{\rm c} dA = \frac{{GM^2 }}{{r^2 }}\,dr,
\]
where $M = \sum {m_i}$ with the sum going over all the particles. Therefore, using equation (\ref{c3}) we can rewrite the previous equation in terms of the velocities as
\[
p_{\rm c} dA = \frac{{M\left\langle {v^2 } \right\rangle }}{r}\,dr.
\]
Now $dA/dr=2A/r$, hence the two-dimensional Curvature Pressure is
\[
p_{\rm c}  = \frac{{M\left\langle {v^2 } \right\rangle }}{{2A}}.
\]
Thus in this two-dimensional model the Curvature Pressure is like the average kinetic energy per unit area. This simple Newtonian model provides a guide as to what the Curvature Pressure would be in the full General Relativistic model.
\par
The extension to different particle masses and velocities uses the basic property of General Relativity that gravitation is an acceleration and not a force. This is supported by \citet*{Eotvos22}, \citet{Dicke64}, and \citet{Braginskii71} who have shown that the passive gravitational mass is equal to the inertial mass to about one part in $10^{12}$. The usual interpretation of this agreement is that they are fundamentally the same thing. However, an alternative viewpoint is that the basic equation is wrong and that the passive gravitational mass and the inertial mass should not appear in Newton's  gravitational equation. Consequently Newton's gravitational equation is an equation of accelerations and not of forces. The equation for Curvature Pressure in a 3 dimensional high temperature plasma is
\begin{equation}
\label{c4}
p_{\rm c}  = \frac{1}{3}\left\langle {\gamma^2 -1}\right\rangle \rho c^2,
\end{equation}
where $\gamma$ is the Lorentz factor and $\langle \rangle$ denotes an average.
\par
In effect, my hypothesis is that the cosmological model must include this Curvature Pressure as well as thermodynamic pressure. Note that although this has a similar form to thermodynamic pressure it is quite different. In particular, it is proportional to an average over the squared velocities and the thermodynamic pressure is proportional to an average over the kinetic energies. This means that, for plasma with free electrons and approximate thermodynamic equilibrium, the electrons will dominate the average due to their much larger velocities.
\par
Including Curvature Pressure into the Friedmann equations provides stable static cosmological model. Including Curvature Pressure from Eq.~\ref{c4}  the modified Friedmann equations are
\begin{eqnarray*}
\label{c5}
\ddot R & = & -\frac{4\pi G\rho}{3} \left[{1 - \left\langle{\gamma^2-1} \right\rangle } \right]R,  \\
\dot R^2 & = & \frac{8\pi G\rho}{3}R^2  - c^2.  \\
\end{eqnarray*}
Clearly, there is a static solution if $<\gamma^2-1>=1$, in which case $\ddot R=0$. The second equation, with $\dot R = 0$  provides the radius of the universe which is given by
\begin{equation}
\label{c6}
R = \sqrt{\frac{3c^2}{8\pi G\rho}} {\mbox{\, }} = \sqrt{\frac{3c^2}
{8\pi GM_{\rm H} N}}.
\end{equation}
Thus, the model is a static cosmology with positive curvature. Although the geometry is similar to the original Einstein static model, this cosmology differs in that it is stable. The basic instability of the static Einstein model is well known \citep{Tolman34,Ellis84}. On the other hand, the stability of Curvature Cosmology is shown by considering a perturbation  $\Delta R$, about the equilibrium position. Then the perturbation equation is
\begin{equation}
\label{c7}
\Delta \ddot R \propto \left(\frac{d\langle\gamma^2-1\rangle}
{dR} \right)\Delta R.
\end{equation}
For any realistic equation of state for the cosmic plasma, the average velocity will decrease as $R$ increases. Thus the right-hand side is negative, showing that the result of a small perturbation is for the universe return to its equilibrium position. Thus, Curvature Cosmology is intrinsically stable. Of theoretical interest is that Eq.~\ref{c7} predicts that oscillations could occur about the equilibrium position.
\subsection{X-ray background radiation}
\label{s8.3}
Since \citet{Giacconi62} observed the X-ray background, there have been many suggestions made to explain its characteristics. Although much of the unresolved X-ray emission comes from active galaxies, there is a part of the spectrum between about 10 keV and 1 MeV that is not adequately explained by emission from discrete sources.
\par
Curvature Cosmology can explain the X-ray emission in the energy range from about 10 keV to 300 keV as coming from a very hot intergalactic plasma.
A simple model has a mixture of hydrogen with 8\% helium and a measured density of $N=1.55\pm0.01$ hydrogen atoms per cubic metre or 2.57$\times10^{-27}\,$kg\,m$^{-3}$.
\par
For this density the predicted a temperature is $2.56\times 10^9\,$K for the cosmic plasma. The temperature estimated from fitting the X-ray data is $(2.62\pm 0.04)\times10^9\,$K which is a good fit. Although this is similar to early explanations of the X-ray emission, it differs in that it depends on the current plasma density. The earlier explanations required the X-ray emission to come from a plasma with about three times that density which conflicted with other observations.
\subsection{Nuclear abundances}
\label{s8.4}
One of the successes of the standard model is in its explanation of the primordial abundances of the light elements.  In Curvature Cosmology, the primordial abundance refers to the abundance in the cosmic plasma from which the galaxies are formed.
\par
The first point to note is that  the predicted temperature of the cosmic plasma is $2.56 \times 10^9 {\mbox{ K}}$
at which temperature nuclear reactions can proceed.  It is postulated that  there is a continuous recycling of material from the cosmic gas to galaxies and stars and then back to the gas. Because of the high temperature, nuclear reactions will take place whereby the more complex nuclei are broken down to hydrogen, deuterium, and helium.
\subsection{Cosmic microwave background radiation }
\label{s8.5}
The Cosmic microwave background radiation (CMBR) is produced by very high energy electrons via Curvature Redshift radiation in the cosmic plasma. With $N=1.55$ the predicted temperature of the CMBR is 3.18\,K to be compared with an observed value of 2.725\,K \citep{Mather90}.  This prediction does depend on the nuclei mix in the cosmic plasma and could vary from this value by several tenths of a degree.
\par
Although the CMBR photons are subject to continuous Curvature Redshift they will be quantised, and since all energy levels are freely available, the black body (Plank function) is their thermal equilibrium spectrum.
\par
Differences in the local environment, especially  high density lower temperature gas clouds, will decrease the flux density of the CMBR  and could explain some of the  observed spatial fluctuations in the CMBR.
\par
\subsection{No Dark matter and  galactic rotation}
\label{s8.6}
In 1937 \citet{Zwicky37} found  in an analysis of the Coma cluster of galaxies that the ratio of total mass obtained by using the virial theorem to the total luminosity  was 500 whereas the expected ratio was 3. The virial theorem  is a statistical theorem that states that for an inverse square law the average kinetic energy of a bound system is equal to half the potential energy. This huge discrepancy was the start of the concept of dark matter. It is surprising that in more than eight decades since that time there is no direct evidence for dark matter. Similarly the concept of dark energy (some prefer quintessence) has been introduced to explain discrepancies in the observations of type 1a supernovae.
\par
X-ray observations show that the Coma cluster has a large plasma cloud in its centre. The Curvature Cosmology model is that the galactic velocity dispersion in the cluster is entirely due to Curvature Redshift of photons passing through the central plasma cloud. For 583 galaxies the rms (root-mean-square) velocity was 893 km\,s$^{-1}$ and the computed theoretical value was $554$ km\,s$^{-1}$. Considering that it was assumed that both the galaxy distribution and plasma distribution had a very simple geometries this shows that Curvature Cosmology can explain the velocity dispersion in the Coma cluster.
\par
One of the most puzzling questions in astronomy is: why  the observed velocity of rotation in spiral galaxies does not go to zero towards the edge of the galaxy. Simple Keplerian mechanics suggest that there should be a rapid rise to a maximum and then a decrease in velocity that is inversely proportional to the square root of the radius once nearly all the mass has been passed. Although the details vary between galaxies, the observations typically show a rapid rise and then an essentially constant tangential velocity as a function of radius out to distances where the velocity cannot be measured due to lack of material. The standard explanation is that this is due to the gravitational attraction of a halo of dark matter  that extends well beyond the galaxy.
\par
Observations show that our own Galaxy and other spiral galaxies have a gas halo that is larger than the main concentration of stars. It is clear that if the observed redshifts are due to Curvature Redshift acting within this halo, the halo must be asymmetric; otherwise, it could not produce the asymmetric rotation curve. Now the observed velocities in the flat part of the curves are typically 100 to 200 km$\;$s$^{-1}$. For realistic values of the densities and sizes of the halo, the velocity is about 163 km$\,$s$^{-1}$. Thus, the magnitude is feasible.
Although there could be a natural asymmetry in a particular galaxy, the fact that the flattened rotation curve is seen for most spiral galaxies suggests that there is a common cause for the asymmetry. One possibility is that the asymmetry could arise from ram pressure due to the galaxy moving through the intergalactic medium.
\par
Although the explanation for galactic rotation observations is limited, Coma cluster observations show no support for dark matter.
Since Curvature Cosmology can explain all the supernova observations there is nor support for dark energy.
\subsection{No Black Holes}
\label{s8.7}
A theory of Curvature Pressure in a very dense medium where quantum mechanics dominate and where general relativity may be required is needed to develop this model. Nevertheless  it is clear that Curvature Pressure  would resist a hot compact object from collapsing to a black hole. Because of the potential energy released during collapse, it is extremely unlikely for a cold object to stay cold long enough to overcome the Curvature Pressure and collapse to a black hole.
\par
What is expected is that the final stage of gravitational collapse is a very dense object, larger than a black hole but smaller than a neutron star. This compact object  would have most of the characteristics of black holes. Such objects could have large masses and be surrounded by accretion discs.  Thus, many of the observations that are thought to show the presence of a black hole could equally show the presence of these compact objects.
\par
If the compact object is rotating there is the tantalising idea that Curvature Pressure  may produce the emission of material in two jets along the spin axis. This could be the ''jet engine" that produces the astrophysical jets seen in stellar-like objects and in many huge radio sources.  Furthermore this could be a mechanism to return material to the cosmic plasma. Currently there are no accepted models for the origin of these jets.
\subsection{Olber's Paradox}
\label{s8.8}
In Curvature Cosmology Olber's Paradox is not a problem.  Visible light from distant galaxies is shifted into the infrared where it is no longer seen and the energy is eventually absorbed back into the cosmic plasma.  Everything is recycled. The plasma radiates energy into the microwave background radiation and into X-rays. The galaxies develop from the cosmic plasma, stars are formed which  pass through their normal evolution. Eventually all their material is returned to the cosmic plasma.
\subsection{Basic equations for Curvature Cosmology}
\label{s8.9}
The geometry is that of a three-dimensional surface of a four-dimensional hyper sphere. For this geometry the  radius is $r=R\chi$ where
\begin{equation}
\label{e12}
\chi = \ln(1+z)/\sqrt{3}.
\end{equation}
\par
(NB. work prior to 2009 has $\chi =\ln(1+z)/\sqrt{2}$))\\
The area is
\begin{equation}
\label{e8}
A(r) = 4\pi R^2 \sin ^2 (\chi ).
\end{equation}
The surface is finite and  $\chi$ can vary from 0 to $\pi$. The volume within a redshift $z$ is given by
\begin{equation}
\label{e9}
V(z) = 2\pi R^3 \left[ {\chi  - \frac{1}{2}\sin (2\chi )}\right]. \nonumber \\
\end{equation}
Using the density $N=1.55\,{m}^{-3})$  the Hubble constant is predicted to be
\begin{eqnarray}
\label{c8}
H & = &-\frac{c}{E}\frac{{dE}}{{ds}} =\left( 8\pi GM_{\rm H} N \right)^{\frac{1}{2}} \nonumber \\
  & = & 51.69 N^{\frac{1}{2}}\;\mbox{kms}^{-1}\;\mbox{Mpc}^{-1} \\
  & = & 64.4\pm0.2\;\mbox{kms}^{-1}\;\mbox{Mpc}^{-1}  \nonumber.
\end{eqnarray}
The only other result required here  is the equation for the distance modulus ($\mu=m-M$),  which is
\begin{equation}
\label{e10}
\mu= 5\log_{10}[(\sqrt{3} \sin(\chi))/h] + 2.5\log_{10}(1+z) + 42.384.\\
\end{equation}
where h=H/(100 $\mbox{kms}^{-1}\;\mbox{Mpc}^{-1}$).
\par

\subsection{Basic consequences of Curvature Cosmology}
\label{s8.10}
Since the ramifications of a static universe are quite profound, a list of the major consequences of Curvature Cosmology is given here. All the numerical results are derived using the cosmic plasma  density $N=1.55$ H atoms m$^{-3}$.
\begin{enumerate}
\item It obeys the perfect cosmological principle
\item It is stable (Section~\ref{s8.2}).
\item There is no dark matter. (Section~\ref{s8.6})
\item There is no dark energy. Meaningless.
\item There is no inflation. Meaningless.
\item There is no horizon problem. Meaningless.
\item The cosmic plasma has a density N$=1.55\pm0.01\,$M$_H\,{m}^{-3})$.
\item The cosmic plasma has a temperature of $(2.64\pm0.04)\times 10^{9}$ K.
\item The Hubble constant is $64.4\pm0.2\;\mbox{kms}^{-1}\;\mbox(Mpc)^{-1}$.
\item It is consistent with supernovae observations.
\item It is consistent with GRB observations.
\item It is consistent with quasar luminosity observations.
\item It is consistent with galaxy luminosity observations.
\item It is consistent with Tolman surface brightness observations.
\item It is consistent with radio source counts.
\item It is consistent with quasar variability
\item It is consistent with angular size observations.
\item It is can explain the Cosmic Microwave Background Radiation (Section~\ref{s8.5}).
\item The CMBR radiation has temperature of $3.18$ K (Section~\ref{s8.5}.
\item It is can provide a partial explanation for fluctuations in CMBR.
\item It is can provide a partial explanation for galactic rotation curves (Section~\ref{s8.6}).
\item It is can explain the X-ray background radiation (Section~\ref{s8.3}).
\item It is can possibly explain the cosmic nuclear abundances (Section~\ref{s8.4}).
\item Curvature redshift can be investigated with laboratory measurements.
\item There are no black holes (Section~\ref{s8.7}).
\item Universal radius: $3.11\times 10^{26}$ m or $1.008\times 10^{10}$ pc.
\item Volume: $8.95\times 10^{80}$ m$^3$ or $2.02\times10^{31}$ pc$^3$.
\item Mass: $2.54\times 10^{54}$ kg or $1.28\times10^{23}$ $\cal M_\bigodot$.
\end{enumerate}
\par
Of interest is that in Curvature Cosmology (CC) distant objects will always have a fainter absolute magnitude than for the standard model. Table~\ref{t6} shows the distance moduli for both cosmologies, their difference, and the absolute flux density ratio for a range redshifts.
\begin{table}[b]
\caption{Relative absolute magnitudes}
\centering
\label{t6}
\begin{tabular}{rrccr}
\hline
Redshift &  $\Lambda$-CDM   & CC    &  diff.& ratio\\ \hline
0     &  42.394     & 42.394     & 0.000 &  1.000 \\
1.0   &  43.512     & 43.067     & 0.445 &  1.507 \\
2.0   &  45.189     & 44.418     & 0.771 &  2.035 \\
5.0   &  47.425     & 45.078     & 1.447 &  3.792 \\
10.0  &  49.102     & 46.927     & 2.175 &  7.412 \\
20.0  &  50.750     & 47.629     & 3.122 &  17.729 \\
50.0  &  52.884     & 48.050     & 4.834 &  85.859 \\
100.0 &  54.468     & 47.682     & 6.786 & 517.962 \\
\hline
\end{tabular}
\end{table}
\section{\bf{CONCLUSION}}
\label{s9}
The first part of this paper argued that the only effect of cosmology on supernovae light curves is to change the scaling parameters of peak flux density and width. The shape of the light curve is intrinsic to the supernovae and is unchanged by cosmology.
\par
Next, it was argued that the redshift of photons is a measure of their energy and could be caused by any systematic energy loss or by time dilation.
\par
In Section~\ref{s4} and \ref{s5} it has been shown that there is a major problem in using SALT2, and similar calibration methods, to remove the intrinsic wavelength dependence of widths from type Ia supernovae light-curve observations. The process of generating the templates means that if the observed light curves have widths that contain the effects of time dilation, these effects are incorporated into the template. The subsequent use of the template will remove this time dilation affects whether artificial or genuine, from the  new observations.
\par
Consequently, SALT2 calibrated light curves cannot contain any cosmological data that is in the form of a power law. Consequently previous analyses of type Ia supernovae gave self-consistent results because of a flaw in the standard analysis program SALT2.
\par
The light curve widths of type Ia supernovae are consistent with no time dilation with an exponent of $0.083\pm0.024$ which is completely inconsistent with standard time dilation which means that the universe is static.
\par
The absolute magnitudes are consistent with Curvature Cosmology with an exponent, $\alpha$,  of $-0.050\pm0.030$, whereas the $\Lambda$-CDM  has an exponent of $-0.014\pm0.030$. From the excellent agreement of the $\Lambda$-CDM  model it is apparent that  the $\Lambda$-CDM distance modulus has been modified to achieve this goal. This has occurred because of the strong belief that  the standard time dilation and  the $\Lambda$-CDM model are both  valid.
\par
One way to partially validate these conclusions would be to redo the SALT2 analysis without the initial division of the epoch differences by $(1+z)$.
\par
In addition the duration of Gamma Ray Bursts are completely consistent with a static universe.
\par
\acknowledgments
\label{s11}
This research has made use of the NASA/IPAC Extragalactic Database (NED) that is operated by the Jet Propulsion Laboratory, California Institute of Technology, under contract with the National Aeronautics and Space Administration. The calculations have used Ubuntu Linux and the graphics have used the DISLIN plotting library provided by the Max-Plank-Institute in Lindau.
\appendix
\section{\bf{SOURCE OF SUPERNOVAE OBSERVATIONS}}
\label{a1}
All of the original type Ia supernovae observations  were retrieved from the SNANA \citep{Kessler09} in the download package $snana.tar.gz$ on the website \url{http://www.snana.uchicago.edu} using the index  files shown in Table~\ref{t7}.
\begin{table}
\caption{Index source files for SNANA data}
\label{t7}
\begin{tabular}{l}
\hline
\hline
   file                                \\ \hline
 { lcmerge/LOWZ\_JRK07}                         \\
 {  lcmerge/JLA2014\_CSP.LIST}                   \\
 {  lcmerge/JLA2014\_CfAIII\_KEPLERCAM.LIST}     \\
 {  lcmerge/SNLS3year\_JRK07.LIST}               \\
 {  lcmerge/SDSS\_allCandidates+BOSS\_HEAD.FITS} \\
 {  lcmerge/JLA2014\_SNLS.LIST}                  \\
 {  lcmerge/JLA2024\_HST.LIST}                   \\
 {  lcmerge/SDSS\_HOLTZ08}                       \\ \hline
\end{tabular}
\end{table}
\par
A current SALT2 template file for the JLA (Joint Light-curve Analysis) analysis was taken from the SNANA  website in the directory $models/SALT2/\-SALT2/JLA\-B14$.
\par
The Pan-STARSS supernovae were accessed from the site \url{https://archive.stsci.edu/prepds/ps1cosmo/jones} and the file datatable.html.
\par
Basic information for all the filters used is shown in Table~\ref{t8}  where column~1 is the filter name, column~2 is the mean wavelength in $\mu\,$m, column~3 (N)  is the final number of supernovae with a valid light curve for this filter, and column~4 is the HST  filter name.
\begin{table}[b]
\caption{Filter characteristics}
\label{t8}
\begin{tabular}{rcrl}
\hline
\hline
Name & Wavelength/$\mu\,m$& N& HST\\ \hline
 B & 0.436 &  202 &\\
 V & 0.541 &  205 \\
 R & 0.619 &  130 &\\
 I & 0.750 &  239&\\
 g & 0.472 & 1,933 &\\
 r & 0.619 & 2,035 &\\
 i & 0.750 & 2,071 &\\
 z & 0.888 & 1,936 &\\
 6 & 0.907 &     5 & F850LP\\
 7 & 1.249 &     1 & F125W\\

\hline
\end{tabular}
\end{table}
\section{\bf{SOURCE OF GRB OBSERVATIONS}}
\label{a2}
The raw GRB data was taken from  \url{$https://swift.gsfc.nasa.gov/archive/grb_table$}
that had burst durations longer than  2 seconds and valid measurements for the redshift, T$_{90}$, the fluence and the peak one-second photon flux rate. The data labelled ``Zhang'' comes from Table~1 in \citet{Zhang13}.
\section{\bf{Equations for $\Lambda$-CDM  COSMOLOGY}}
\label{sa2}
The equations needed for the modified $\Lambda$-CDM model \citep{Hogg99,  Goliath01, Barboza08}, with $\Omega_M=0.27$,
$\Omega_K=0$  and where $h$ is the reduced Hubble constant, are listed below. The symbol $w^*$ is used for the acceleration parameter in order to avoid confusion with the width $w$. These equations depend on the function $E(z)$ defined here by
\begin{equation}
\label{eeb1}
E(z) = \int_0^z \frac{dz}{\sqrt{\Omega_M (1+z)^3+(1-\Omega_M)(1+z)^{(1+w^*)}}}.
\end{equation}
The distance modulus is
\begin{equation}
\label{eb2}
\mu_B(z)=5\log_{10}(E(z)(1+z)/h)+ 42.384.
\end{equation}
The co-moving volume is
\begin{equation}
\label{eb3}
v_B(z)=\frac{4\pi}{3}(2.998E(z)/h)^3 {  Gpc}^3.
\end{equation}
The equation of state parameter $w^*$ in the expansion model distance modulus is included to investigate the effects of including the cosmological constant.  \citet{Conley11} found that  the  parameter, $w^*$,  has a value $w^*=-0.91$, whereas \citet{Sullivan11} found that $w^*=-1.069$.   Although its actual value is not  critical for this paper the value of $w^*$ is chosen to be  $w^*=-1.11$, so that $E_B$ would be the best fiducial constant with the values for the  peak magnitudes and stretch factors provided by B14.
\label{lastpage}

\end{document}